\documentclass[10pt,twoside]{article}
\usepackage{amssymb,amsmath}
\def\volnum{Manuscrit}

\usepackage[ansinew]{inputenc}
\usepackage{typearea,epsfig,rotating,times}

\def\a{\alpha}
\def\b{\beta}
\def\g{\gamma}
\def\d{\delta}
\def\e{\epsilon}

\def\vt{\vartheta}

\usepackage{aflbcours}
\pdebut{1}
\def\pacs#1{\LP P.A.C.S.: #1}
\title{\'Elie Cartan's torsion in geometry and in
  field theory, an essay}
\titleshort{Torsion in geometry and in
  field theory}
\author{Friedrich W.\ Hehl$^1$ and Yuri N.\ Obukhov$^2$\\
\begin{footnotesize}{\rm email:
  hehl@thp.uni-koeln.de, yo@thp.uni-koeln.de}
\end{footnotesize}}
\authorshort{F.W.\ Hehl and Yu.N.\ Obukhov}
\address{Institute for
  Theoretical Physics, University of Cologne\\ 50923 K\"oln, Germany\\  
$^1$Dept.\ of Physics \& Astronomy, Univ. of Missouri-Columbia\\ Columbia, MO
  65211, USA\\ $^2$Dept.\ of Theoretical Physics, Moscow 
State University\\ 117234 Moscow, Russia}

\begin{document}
\maketitle

\vskip 1cm
\begin{abstract}
ABSTRACT. 

We review the application of torsion in field theory.  First we show
how the notion of torsion emerges in differential geometry. In the
context of a Cartan circuit, {\it torsion} is related to {\it
  translations} similar as curvature to rotations. Cartan's
investigations started by analyzing Einsteins general relativity
theory and by taking recourse to the theory of Cosserat continua. In
these continua, the points of which carry independent translational
and rotational degrees of freedom, there occur, besides ordinary
(force) stresses, additionally {\it spin moment stresses.} In a
3-dimensional continuized crystal with dislocation lines, a linear
connection can be introduced that takes the crystal lattice structure
as a basis for parallelism. Such a continuum has similar properties as
a Cosserat continuum, and the dislocation density is equal to the
torsion of this connection. Subsequently, these ideas are applied to
4-dimensional spacetime. A translational gauge theory of gravity is
displayed (in a Weitzenb\"ock or teleparallel spacetime) as well as
the viable Einstein-Cartan theory (in a Riemann-Cartan spacetime). In
both theories, the notion of torsion is contained in an essential way.
Cartan's spiral staircase is described as a 3-dimensional Euclidean
model for a space with torsion, and eventually some controversial
points are discussed regarding the meaning of torsion. {\it file
  deBroglie10.tex, 09 Nov 2007}
\\

R\'ESUM\'E.

\end{abstract}

\section{A connection induces torsion and curvature}

{\em ``...the essential achievement of general relativity, namely to overcome
`rigid' space (ie the inertial frame), is {\em only indirectly} connected
with the introduction of a Riemannian metric. The directly relevant
conceptual element is the `displacement field' ($\Gamma^l_{ik}$), which
$\,$expresses the$\,$ infinitesimal displacement of vectors. It is this which
replaces the parallelism of spatially arbitrarily separated vectors fixed
by the inertial frame (ie the equality of corresponding components) by an
infinitesimal operation. This makes it possible to construct tensors by
differentiation and hence to dispense with the introduction of `rigid'
space (the inertial frame). In the face of this, it seems to be of
secondary importance in some sense that some particular $\Gamma$ field can
be deduced {}from a Riemannian metric...''} 

\hfill A.\ Einstein (4 April 1955)\footnote{Preface in `Cinquant'anni
  di Relativit\`a 1905--1955.' M.\ Pantaleo, ed.. Edizioni Giuntine
  and Sansoni Editore, Firenze 1955 (transaltion {}from the German
  original by F.\ Gronwald, D.\ Hartley, and F.W.\ Hehl). For the role
  that generalized connections play in physics, see Mangiarotti and
  Sardanashvily \cite{Sardan}.}  \bigskip

\begin{figure}\label{dislFig0}
\begin{center}
\includegraphics[width=10cm]{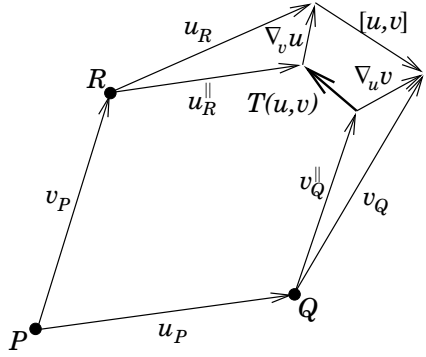}
\caption{{\em On the geometrical interpretation of torsion,} see
  \cite{birkbook}: Two vector fields $u$ and $v$ are given. At a point
  $P$, we transport parallelly $u$ and $v$ along $v$ or $u$,
  respectively. They become $u_{\rm R}^{||}$ and $v_{\rm Q}^{||}$. If
  a torsion is present, they don't close, that is, a {\it closure
    failure} $T(u,v)$ emerges. This is a schematic view. Note that the
  points $R$ and $Q$ are infinitesimally near to $P$. A proof can be
  found in Schouten \cite{Schouten1954}, p.127.}
\end{center}
\end{figure}

On a differential manifold, we can introduce a linear connection, the
components of which are denoted by $\Gamma_{ij}{}^k$. The connection
allows a parallel displacement of tensors and, in particular, of
vectors, on the manifold. We denote (holonomic) coordinate indices
with Latin letters $i,j,k,\dots =0,1,2,\dots,n-1$, where $n$ is the
dimension of the manifold. A vector $u=u^k\partial_k$, if parallelly
displaced along $dx^i$, changes according to
\begin{equation}\label{connection}
\delta^{||}u^k=-\Gamma_{ij}{}^k u^j dx^i \,.
\end{equation}
Based on this formula, it is straightforward to show that a
non-vanishing Cartan torsion,\footnote{According to Kiehn
  \cite{Kiehn}, one can distinguish at least five different notions of
  torsion. In our article, we treat Cartan's torsion of 1922, as it is
  established in the meantime in differential geometry, see Frankel
  \cite{Frankel}, p.245. We find it disturbing to use the same name
  for different geometrical objects.}
\begin{equation}\label{torsionhol}
T_{ij}{}^k=\Gamma_{ij}{}^k-\Gamma_{ji}{}^k\equiv
2\Gamma_{[ij]}{}^k\ne 0\,,
\end{equation}
breaks infinitesimal parallelograms on the manifold, see
Fig.\ref{dislFig0}. Here for antisymmetrization we use the
abbreviation $[ij]:=\frac 12\left(ij-ji \right)$ and for
symmetrization $(ij):=\frac 12\left(ij+ji \right)$, see
\cite{Schouten1954}. There emerges a {\it closure failure}, i.e., a
parallelogram is only closed up to a small translation.

In GR, the connection is identified with the Christoffel symbol
$\Gamma_{ij}{}^k=\{_{i}{}^{k}{}_{j}\}$ and is as such symmetric
$\{_{i}{}^{k}{}_{j}\}=\{_{j}{}^{k}{}_{i}\}$. In other words, the
torsion vanishes in GR.

The torsion surfaces more naturally in a frame formalism.  At each
point we have a basis of $n$ linearly independent vectors
$e_\a=e^i{}_\a\partial_i$ and the dual basis of covectors
$\vt^\b=e_j{}^\b dx^j $, the so-called coframe, with $e_\a\rfloor
\vt^\b=\d_\a^\b$ (the interior product is denoted by $\rfloor$). We
denote (anholonomic) frame indices with Greek letters $\a,\b,\g,\dots
=0,1,2,\dots,n-1$. The connection is then introduced as
1-form\footnote{The relation between $\Gamma_{i\a}{}^\b$ and the
  holonomic $\Gamma_{ij}{}^k$ in (\ref{connection}) is
  $\Gamma_{i\a}{}^\b=e^j{}_\a
  e_k{}^\b\Gamma_{ij}{}^k+e^j{}_\a\partial_i e_j{}^\b $.}
$\Gamma_\a{}^\b= \Gamma_{i\a}{}^\b dx^i$, and, for a form $w^A$, we
can define a covariant exterior derivative according to $Dw^A := dw^A
+ \rho_B{}^{A\,\alpha}{}_\beta\,\Gamma_\alpha{}^\beta\wedge w^B $.
Here the coefficients $\rho_B{}^{A\,\alpha}{}_\beta$ describe the
behavior of $w^A$ under linear transformations, for details see
\cite{TrautmanSUNY} and \cite{birkbook}, p.199, and $\wedge$ denotes
the exterior product. Then the torsion 2-form is defined as
\begin{equation}\label{torsion}
T^\a:= D\vt^\a=d\vt^\a+\Gamma_\b{}^\a\wedge \vt^\b\,.
\end{equation}
If the frames are chosen as coordinate frames, then $d\vt^\a=0$ and
the definition (\ref{torsion}) degenerates to (\ref{torsionhol}).
{}From (\ref{torsion}) we can read off that $T^\a$ is a kind of a
field strength belonging to the `potential' $\vt^\a$.

Since we introduced a connection $\Gamma_\a{}^\b$, we can define in
the conventional way the RC-curvature,
\begin{equation}\label{curvature}
R_\a{}^\b:=d\Gamma_\a{}^\b+\Gamma_\g{}^\b\wedge\Gamma_\a{}^\g\,.
\end{equation}
If we differentiate (\ref{torsion}) and (\ref{curvature}), we find
straightforwardly the first and the second Bianchi identities,
respectively,\footnote{In 3 dimensions we have $1\times(3+3)=6$ and in
  4 dimensions $4\times(4+6)=40$ independent components of the Bianchi
  identities.}
\begin{equation}\label{Bianchi}
DT^\a= R_\b{}^\a\wedge \vt^\b\,, \qquad DR_\a{}^\b=0\,.
\end{equation}

We can recognize already here, how closely torsion and curvature are
interrelated. Moreover, it is clear, that torsion as well as well as
curvature are notions linked to the process of parallel displacement
on a manifold and are as such something very particular.

\section{Cartan circuit: Translational and rotational misfits}

Since in the applications we have in mind the metric plays an
essential role, we will now introduce --- even though it is not
necessary at this stage --- besides the connection $\Gamma_\a{}^\b $,
a (symmetric) metric $g_{ij}=g_{ji}$ that determines distances and
angles. The line element is given by
\begin{equation}\label{metric}
  ds^2=g_{ij}dx^i\otimes dx^j=g_{\a\b}\vt^\a\otimes\vt^\b\,.
\end{equation}
We assume that the connection is compatible with the metric, i.e., the
nonmetricity $Q_{\a\b}$ vanishes:
\begin{equation}\label{nonmetricity}
Q_{\a\b}:=-Dg_{\a\b}=0\,.
\end{equation}
A space fulfilling this condition is called a {\it Riemann-Cartan\/}
(RC) {\it space}. We can solve (\ref{nonmetricity}) with respect to
the symmetric part of the (anholonomic) connection:
\begin{equation}\label{symmconn}
\Gamma_{(\a\b)}=\frac 12d g_{\a\b}\,.
\end{equation}

Furthermore, we will choose an {\it orthonormal coframe}.
This can be done in any dimension $n>1$. We will apply the formalism
to the 4-dimensional (4D) spacetime with Lorentzian metric
$g_{\a\b}={\rm diag}(-1,1,1,1)$ or to the 3D space with Euclidean
metric $g_{\a\b}={\rm diag}(1,1,1)$. Then, due to (\ref{symmconn}), we
find a vanishing symmetric part of the anholonomic connection.
Accordingly, we have in a RC-space as geometrical field variables the
orthonormal coframe $\vt^\a=e_i{}^\a dx^i$ and the metric-compatible
connection $\Gamma^{\a\b}=\Gamma_i{}^{\a\b}dx^i= -\Gamma^{\b\a}$.

Now we are prepared to characterize a RC-space in the way Cartan did
it. Locally a RC-space looks Euclidean, since for any single point
$P$, there exist coordinates $ x^{i}$ and an orthonormal coframe
$\vt^{\alpha}$ in a neighborhood of $P$ such that
\begin{equation}\left\{\begin{array}{rcl} \vt^{\alpha}&=&\,\delta
      ^{\alpha}_{i}dx^{i}\\ \Gamma
      _{\alpha}{}^{\beta}&=&\,0
         \end{array}
       \right\}\quad {\rm at}\,\,\, P\,, \label{propo}
\end{equation} 
where $\Gamma_{\alpha}{}^{\beta}$ are the connection 1--forms referred
to the coframe $\vt^{\alpha}$, see Hartley \cite{Hartley} for details.
Eq.(\ref{propo}) represents, in a RC-space, the anholonomic analogue
of the (holonomic) Riemannian normal coordinates of a Riemannian
space.

Often it is argued incorrectly that is RC-space normal frames cannot
exist, since torsion, as a tensor, cannot be transformed to zero. In
this context it is tacitly assumed that the starting point are
Riemannian normal coordinates and the torsion is `superimposed'.
However, since only a {\it natural}, i.e., a holonomic or coordinate
frame is attached to Riemannian normal coordinates, one is too
restrictive in the discussion right {}from the beginning. And, of
course, the curvature is also of tensorial nature -- and still
Riemannian normal coordinates do exist.

How can a local observer at a point P with coordinates $x^i\ $ tell
whether his or her space carries torsion and/or curvature? The local
observer defines a small loop (or a circuit) originating {}from P and
leading back to P. Then he/she {\it rolls} the local reference space
{\it without sliding} --- this is called Cartan displacement --- along
the loop and adds up successively the small relative translations and
rotations, see Cartan \cite{Cartan1986,Cartan2001}, Schouten
\cite{Schouten1954}, Sharpe \cite{Sharpe} or, for a modern
application, Wise \cite{Wise}. As a computation shows, the added up
{\it translation} is a measure for the {\it torsion} and the {\it
  rotation} for the {\it curvature}. Since the loop encircles a small
2-dimensional area element, Cartan's prescription attaches to an area
element a small translation and a small rotation. Thus, torsion $T^\a$
and curvature $R^{\a\b}=-R^{\b\a}$ are both 2-forms in any dimensions
$n>1$, the torsion is vector-valued, because of the translation
vector, the curvature bivector-valued, because of the rotations.

In this way Cartan visualized a RC-space as consisting of a collection
of small Euclidean granules that are translated and rotated with
respect to each other. Intuitively it is clear that this procedure of
Cartan is similar to what one does in gauge field theory: A rigid (or
global) symmetry, here the corresponding Euclidean motions of
translation and rotations, is extended to a local symmetry. In
four-dimensional spacetime it is the Poincar\'e (or inhomogeneous
Lorentz) group of Minkowski space that is gauged and that yields a
RC-spacetime, see \cite{Jim,Milutin,Erice}.

There are two degenerate cases: A RC-space with vanishing torsion is
the conventional Riemannian space, a RC-space with vanishing
RC-curvature is called a {\it Weitzenb\"ock space}
\cite{Weitzenboeck}, or a space with teleparallelism. We will come
back to this notion later.

We can now list the number of the components of the
different geometrical quantities in a RC-space of 3 or 4 dimensions.
These numbers are reflecting the $3+3$ generators of the 3D Euclidean
group and of the $4+6$ generators of the 4D Poincar\'e group:\medskip

\begin{center}
\begin{tabular}{|c|c|c|c|c|}\hline
  &orthon.\ cofr.&RC-connection&Cartan's torsion&RC-curvature\\
  &$\vt^\a$&$\Gamma^{\a\b}$&$T^\a$&$R^{\a\b}$\\ \hline\hline
  $n=3$&$\;9=3\times 3$&$\;9=3\times 3$&$\;9=3\times 3$&
$\;9=3\times 3$\\ \hline
  $n=4$&$16=4\times 4$&$24=6\times 4$&$24=6\times 4$&$36=6\times 6$\\ \hline
\end{tabular}\medskip
\end{center}

The results of Secs.1 and 2 can all be proven rigorously. They are all
consequences of the introduction of a connection $\Gamma_{ij}{}^k$ and
a metric $g_{ij}$. Let us now turn to a new ideas that influenced
Cartan's thinking in the context of RC-geometry.

\begin{figure}\label{CossFig1}
\begin{center}
\includegraphics[width=7cm]{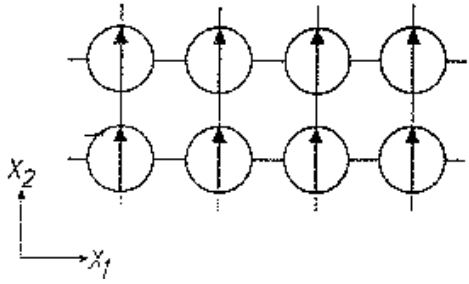}
\end{center}
\caption{Schematic view on a two-dimensional Cosserat continuum:
  Undeformed initial state.}
\end{figure}

\section{The Cosserat continuum}

Cartan, according to his acknowledgment in \cite{Cartan1922}, was
inspired by the brothers Cosserat \cite{Cosserat} and their theory of
a new type of continuum. The classical continuum of elasticity and
fluid dynamics consists of unstructured points, and the displacement
vector $u_i$ is the only quantity necessary for specifying the
deformation. The Cosserats conceived a specific {\it medium with
  microstructure,} see \cite{Guenther,Capriz,GronHehl} and for a
historical review \cite{Badur}, consisting of structured points such
that, in addition to the displacement field $u_i$, it is possible to
measure the rotation of such a structured point by the bivector field
$\omega_{ij}=-\omega_{ji}$, see Fig.2 for a schematic view.

The deformation measures {\it distortion\/} $\b$ and {\it
  contortion\/} $\kappa$ of a linear Cosserat continuum are
($\nabla_i$ is the covariant derivative operator of the Euclidean 3D
space)
\begin{eqnarray}\label{Coss1}
  \b_{ij}\,&=&\,\nabla_i u_j-\omega_{ij}\,,\qquad\omega_{ij}=
-\omega_{ji}\,,\\ \label{Coss2}
  \kappa_{ijk}\,&=&\,\nabla_i\omega_{jk}=-\kappa_{ikj}\,,
\end{eqnarray}
see G\"unter \cite{Guenther} and Schaefer \cite{SchaeferZAMM}. A
rigorous derivation of these deformation measures is given in the
Appendix. In classical elasticity, the only deformation measure is the
strain $\varepsilon_{ij}:=\frac 12(\b_{ij}+\b_{ji})\equiv
\b_{(ij)}=\nabla_{(i}u_{j)}$. Let us visualize these deformations. If
the displacement field $u_1\sim x$ and the rotation field
$\omega_{ij}=0$, we find $\b_{11}=\varepsilon_{11}=const$ and
$\kappa_{ijk}=0$, see Fig.3. This homogeneous strain is created by
ordinary force stresses. In contrast, if we put $u_i=0$ and
$\omega_{12}\sim x$, then $\b_{12}=\omega_{12}\sim x$ and
$\kappa_{112}\sim const$, see Fig.4. This homogeneous contortion is
induced by applied spin moment stresses. Fig.5 depicts the pure
constant antisymmetric stress with $\omega_{12}=const$ and Fig.6 the
conventional rotation of the particles according to ordinary
elasticity. This has to be distinguished carefully {}from the situation
in Fig.4.

\begin{figure}\label{CossFig2}
\begin{center}
\includegraphics[width=11cm]{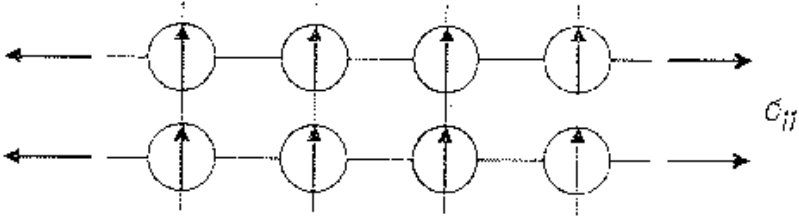}
\end{center}
\caption{Conventional homogeneous strain $\varepsilon_{11}$ of a Cosserat
  continuum: Distance changes of the ``particles'' caused by force
  stress $\sigma_{11}$.}
\end{figure}

\begin{figure}\label{CossFig3}
\begin{center}
\includegraphics[width=8cm]{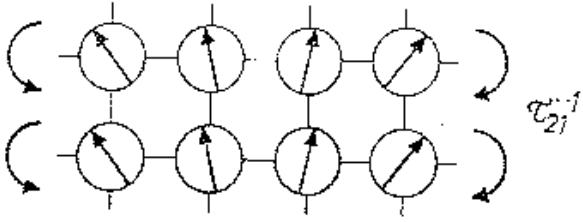}
\end{center}
\caption{Homogeneous contortion $\kappa_{112}$ of a Cosserat continuum:
  Orientation changes of the ``particles'' caused by spin moment
  stress $\tau_{21}{}^1$.}
\end{figure}

Apparently, in addition to the force stress
$\overline{\Sigma}_{ij}\sim \d{\cal H}/\d \b_{ij}$ (here $\cal H$ is
an elastic potential), which is asymmetric in a Cosserat continuum,
i.e., $\overline{\Sigma}_{ij}\ne \overline{\Sigma}_{ji}$, we have as
new response the {\it spin moment} stress
$\overline{\tau}_{ijk}\sim\d{\cal H}/\d \kappa_{kji}$. Hence (force)
stress $\overline{\Sigma}_{ij}$ and {\it spin moment} stress
$\overline{\tau}_{ijk}$ characterize a Cosserat continuum {}from the
static side. We used the overlines for denoting stress and spin moment
stress specifically in 3D.

Only in 3D, a rotation can be described by a vector according to
$\omega^i=\frac 12\epsilon^{ijk} \omega_{jk}$, where
$\epsilon_{ijk}=0,+1,-1$ is the totally antisymmetric 3D permutation
symbol. We chose here the bivector description such that the
discussion becomes independent of the dimension of the continuum
considered. Even though there exist 1D Cosserat continua (wires and
beams) and 2D ones (plates and shells), we will concentrate here,
exactly as Cartan did, on 3D Cosserat continua.

The equilibrium conditions for forces and moments read\footnote{In
  exterior calculus we have $D{\Sigma}_\a+f_\a=0$ and
  $D{\tau}_{\a\b}+\vt_{[\a}\wedge{\Sigma}_{\b]}
  +m_{\a\b}=0$.  These relations are valid in all dimensions $n\ge 1$,
  see \cite{GronHehl}. In 3 dimensions we have $3+3$ and in 4
  dimensions $4+6$ independent components of the ``equilibrium''
  conditions.}
\begin{eqnarray}\label{equilibrium}
  \nabla_j\overline{\Sigma}_i{}^j+f_i=0,\qquad
  \nabla_k\overline{\tau}_{ij}{}^k-\overline{\Sigma}_{[ij]}+m_{ij}=0,
\end{eqnarray}
where $f_i$ are the volume forces and $m_{ij}=-m_{ji}$ volume moments.
They correspond to translational and rotational Noether identities. In
classical elasticity and in fluid dynamics,
$\overline{\tau}_{ij}{}^k=0$ and $m_{ij}=0$; thus, the stress is
symmetric, $\overline{\Sigma}_{[ij]}=0$, and then denoted by
$\overline{\sigma}_{ij}$; for early investigations on asymmetric
stress and energy-momentum tensors, see Costa de Beauregard
\cite{Costa}.

Nowadays the Cosserat continuum finds many applications. As one
example we may mention the work of Zeghadi et al. \cite{grains} who
take the grains of a metallic polycrystal as (structured) Cosserat
particles and develop a linear Cosserat theory with the constitutive
laws $\overline{\Sigma}_{ij}\sim \b_{ij}$ and
$\overline{\tau}_{ijk}\sim \kappa_{kji}$.

\begin{figure}\label{CossFig4}
\begin{center}
\includegraphics[width=6cm]{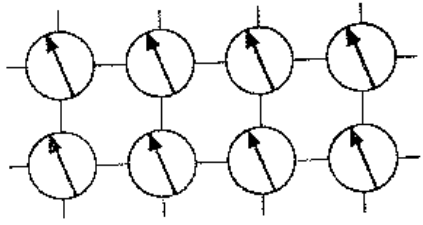}
\end{center}
\caption{Homogeneous Cosserat rotation $\omega_{12}$ of the
  ``particles'' of a Cosserat continuum caused by the antisymmetric
  piece of the stress $\Sigma_{[12]}$.}
\end{figure}

The Riemannian space is the analogue of the body of classical
continuum theory: points and their relative distances is all what is
needed to describe it geometrically; the analogue of the strain
$\varepsilon_{ij}$ of classical elasticity is the metric tensor
$g_{ij}$ of the Riemannian space. In GR, a symmetric ``stress''
$\sigma_{ij}=\sigma_{ji}$ is the response to a variation of the metric
$g_{ij}$.

A RC-space can be realized by a generalized Cosserat continuum.  The
``deformation measures'' $\vt^\a=e_i{}^\a dx^i$ and
$\Gamma^{\a\b}=\Gamma_i{}^{\a\b}dx^i=-\Gamma^{\b\a}$ of a RC-space
correspond to those of a Cosserat continuum:\footnote{This can be seen
  {}from the response of the coframe $e_i{}^\a$ and the Lorentz
  connection $\Gamma_i{}^{\a\b}$ in a RC-space to a local Poincar\'e
  gauge transformation consisting of small translations $\epsilon^\a$
  and small Lorentz transformations $\omega^{\a\b}$,
\begin{eqnarray}\label{PG1}
  \d e_i{}^\a &=&\, -D_i\e^a+e_i{}^\g\omega_\g{}^\a-\e^\g T_{\g i}{}^\a\,,\\
  \d\Gamma_i{}^{\a\b}&=&\,   -D_i\omega^{\a\b}
-\e^\g R_{\g i}{}^{\a\b}\,,\label{PG2}
\end{eqnarray}
see \cite{RMP}, Eqs.(4.33),(4.32); here $D_i:=\partial_i\rfloor D$ are
the components of the exterior covariant derivative.  The second term
on the right-hand-side of (\ref{PG1}) is due to the semi-direct
product structure of the Poincar\'e group. If we put torsion and
curvature to zero, these formulas are analogous to
(\ref{Coss1}),(\ref{Coss2}).}
\begin{equation}\label{map}
  e_i{}^\a\rightarrow \b_{ij}\,,\qquad\Gamma_i{}^{\a\b}
  \rightarrow \kappa_{ijk}\,.
\end{equation}
However, the coframe $\vt^\a$ and the connection $\Gamma^{\a\b}$
cannot be derived {}from a displacement field $u_i$ and a rotation field
$\omega_{ij}$, as in (\ref{Coss1}),(\ref{Coss2}). Such a generalized
Cosserat continuum is called incompatible, since the deformation
measures $\b_{ij}$ and $\kappa_{ijk}$ don't fulfill the so-called
compatibility conditions
\begin{equation}\label{compat}
  \nabla_{[i}\b_{j]k}+\kappa_{[ij]k}=0\,,\qquad
  \nabla_{[i}\kappa_{j]kl}=0 \,,
\end{equation}
see G\"unther \cite{Guenther} and Schaefer
\cite{SchaeferZAMM,Schaefer}. They guarantee that the ``potentials''
$u_i$ and $\omega_{ij}$ can be introduced in the way as it is done in
(\ref{Coss1}),(\ref{Coss2}). Still, also in the RC-space, as {\it
  incompatible\/} Cosserat continuum, we have, besides the force
stress $\overline{\Sigma}_\a{}^i\sim\d{\cal H}/\d e_{i}{}^\a$, the
spin moment stress $\overline{\tau}_{\a\b}{}^i\sim\d{\cal H}/\d
\Gamma_{i}{}^{\a\b}$.  And in the geometro-physical interpretation of
the structures of the RC-space, Cartan apparently made use of these
results of the brothers Cosserat.

In 4D, the stress $\overline{\Sigma}_\a{}^i $ corresponds to
energy-momentum\footnote{This is well-known {}from classical
  electrodynamics: The 3D Maxwell stress generalizes, in 4D, to the
  energy-momentum tensor of the electromagnetic field, see
  \cite{birkbook}.} ${\Sigma}_\a{}^i$ and the spin moment stress
$\overline{\tau}_{\a\b}{}^i$ to spin angular momentum
${\tau}_{\a\b}{}^i$. Accordingly, Cartan enriched the Riemannian space
of GR geometrically by the {\it torsion} $T_{ij}{}^\a$ and statically
(or dynamically) by the {\it spin angular momentum} $\tau_{\a\b}{}^i$
of matter.

\begin{figure}\label{CossFig5}
\begin{center}
\includegraphics[width=7cm]{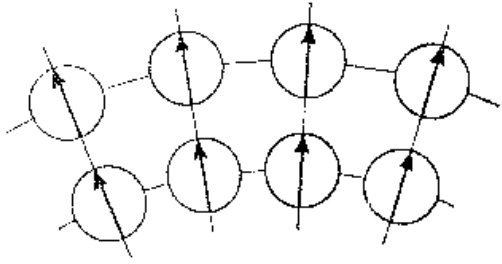}
\end{center}
\caption{Conventional rotation $\partial_{[1}u_{2]}$ of the
  ``particles'' of a Cosserat continuum caused by an inhomogeneous
  strain.}
\end{figure}

\begin{figure}\label{dislFig1}
\begin{center}
\includegraphics[width=7cm]{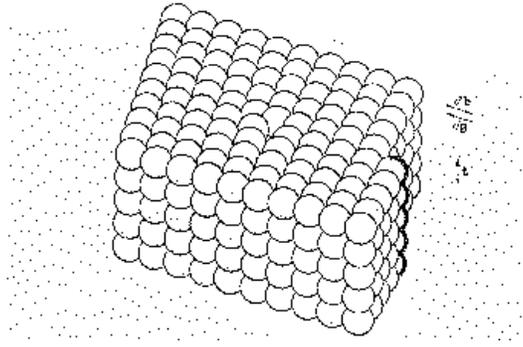}
\end{center}
\caption{{\em Edge dislocation} after Kr\"oner \cite{Kroener1958}: The
  dislocation line is parallel to the vector $\mathbf{t}$. The Burgers
  vector $\d\mathbf{b}$, characterizing the missing half-plane, is
  perpendicular to $\mathbf{t}$. The vector $\d\mathbf{g}$
  characterizes the gliding of the dislocation as it enters the ideal
  crystal.}
\end{figure}

\section{A rule in three dimensions: Dislocation density equals
  torsion}

In the 1930s, the concept of a crystal dislocation was introduced in
order to understand the plastic deformation of crystalline solids, as,
for instance, of iron. Dislocations are one-dimensional lattice
defects. Basically, there exist two types of dislocations, edge and
screw dislocation, see Weertman \& Weertman \cite{Weertman}. In
Fig.7
, we depicted a three-dimensional view on such an edge dislocation in
a cubic primitive crystal. We recognize that one atomic half-plane has
been moved to the right-hand-side of the crystal. The missing
half-plane is characterized by the Burgers vector that is
perpendicular to the dislocation line. The screw dislocation of
Fig.8
\, has again a Burgers vector, but in this case it is parallel to the
dislocation line. In the framework of classical elasticity, at the
beginning of the last century, theories of the elastic field of
singular defect lines had been developed by Volterra, Somigliana, and
others, see Nabarro \cite{Nabarro} and Puntigam \& Soleng
\cite{Puntigam}. These theories could be used to compute the far-field
of a crystal dislocation successfully. For more recent developments in
this field, one may quote Malyshev \cite{Malyshev}, who went beyond
the linear approximation.

\begin{figure}\label{dislFig2}
\begin{center}
\includegraphics [width=7cm]{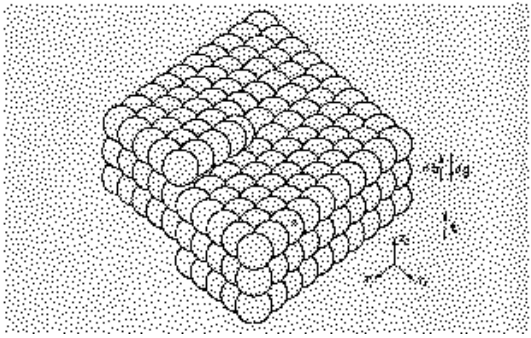}
\end{center}
\caption{{\em Screw dislocation} after Kr\"oner \cite{Kroener1958}: Here
the Burgers vectors is parallel to $\mathbf{t}$.}
\end{figure}

If sufficiently many dislocations populate a crystal, then a continuum
or field theory of dislocations is appropriate, see Kr\"oner's theory
of a continuized crystal \cite{Kroener1986}. In order to give an idea
of such an approach, let us look at a cubic crystal in which several
dislocations are present, see Fig.11
. By averaging over, we can define a dislocation density tensor
$\a_{ij}{}^k=-\a_{ji}{}^k$. The indices $ij$ denote the area element,
here the 12-plane, and $k$ the direction of the Burgers vector, here
only the component $\d b^1$.  Thus, in Fig.11
, only the $\a_{12}{}^1=-\a_{21}{}^1$ components are nonvanishing.

Already in 1953, Nye \cite{Nye1953} was able to derive a relation
between the dislocation density $\a_{ijk}$ and the contortion tensor
$K_{ijk}$, which describes the relative rotations between neighboring
lattice planes:
\begin{equation}\label{contortion}
K_{ijk}=-\a_{ijk}+\a_{jki}-\a_{kij}=-K_{ikj}\,.
\end{equation}
On purpose we took here the letter $K$ similar to the contortional
measure $\kappa$ of a Cosserat continuum, see (\ref{Coss2}). In
Fig.11
, according to Eq.(\ref{contortion}), only $K_{121}=-K_{211}\ne 0$: We
have rotations in the 12-plane if we go along the $x_1$-direction.

At the same time it becomes clear that, {}from a macroscopic, i.e.,
continuum theoretical view, the response of the crystal to its
contortion induced by the dislocations are spin moment stresses
$\tau_{ij}{}^k$, as indicated in Fig.11, see \cite{HK}. This is the
new type of spin moment stress that already surfaced in the Cosserat
continuum in Fig.4.  It is obvious, if one enriches in geometry the curvature
by the {\it torsion,} then in the dynamical side one should allow,
besides stress (in 4D energy-momentum), {\it spin moment stress} (in
4D spin angular momentum).

The ideal reference crystal, in the sense of Cartan, is the undeformed
crystal of Fig.9
. One can imagine to roll it along the dislocated crystal in
Fig.11
.  Then the closure failure of Fig.11 
is determined, provided we define the connection with respect to the
lattice vectors.  In dislocation theory, this is known as the
Frank-Burgers circuit, the closure failure as the Burgers vector. The
cracking of a small parallelogram, defined in the undeformed crystal
in Fig.9, can be recognized in Fig.11. Clearly, this procedure is
isomorphic to the Cartan circuit, as has been proven by Kondo (1952)
\cite{Kondo1952}, Bilby et al.\ \cite{Bilby} and Kr\"oner
\cite{Kroener1958,Kroener1980}.  Thus, it is an established fact that
{\it dislocation density} and {\it torsion} in three dimensions can be
used synonymously.

\begin{figure}\label{dislFig4}
\begin{center}
\includegraphics[width=9cm]{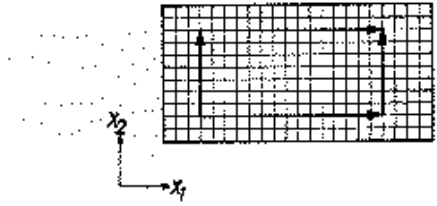}
\end{center}
\caption{{\em The ideal cubic crystal in the undeformed state,}
see \cite{HvdH}: A ``small'' parallelogram has been drawn.}
\end{figure}

We recognize that at each point in a crystal with dislocations a
lattice direction is well-defined, see Fig.11. In other words, a
global teleparallelism is provided thereby reducing the RC-space to a
Weitzenb\"ock space with vanishing RC-curvature, see Fig.12. It can be
shown \cite{Schroedinger} that the connection of a Weitzenb\"ock space
can always be represented in terms of the components of the frame
$e_\a=e^k{}_\a\partial_k$ and the coframe $\vt^\a=e_j{}^\a dx^j$ as
\begin{equation}
\Gamma_{ij}{}^k = e^k{}_\alpha\,\partial_i e_j{}^\alpha\,.\label{Gweitz}
\end{equation}
Accordingly, on the one hand a dislocated crystal carries a torsion
(that is, a dislocation density), on the other hand it provides a
teleparallelism or defines a Weitzenb\"ock space (that is, a space
with vanishing RC-curvature), see, e.g., the discussion of Kr\"oner
\cite{Kroener1993}.

\section{Translation gauge theory of continuously distributed
  dislocations}

What are then the deformational measures in the field theory of
dislocations, see Kr\"oner
\cite{Kroener1958,Kroener1980,Kroener1986,Kroener1993}?  Clearly,
torsion $\a$ or contortion $K$ must be one measure, but what about the
distortion? We turn to the fundamental work of Lazar
\cite{Lazar,LazarA}, Katanaev \cite{Katanaev}, and Malyshev
\cite{Malyshev} on the 3D translational gauge approach to dislocation
theory.  The underlying geometrical structure of the theory is the
affine tangent bundle $A(M)$ over the 3-dimensional base space $M$. It
arises when one replaces at every point of $M$ the usual tangent space
by an affine tangent space. In the affine space, one can perform
translations of the points and vectors, and in this way one the
translation group $T_3$ is realized as an {\it internal symmetry}.

\begin{figure}\label{dislFig5}
\begin{center}
\includegraphics[width=9cm]{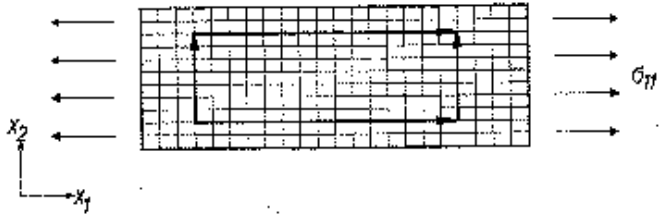}
\end{center}
\caption{Homogeneously strained crystal caused by force stress
  $\sigma_{11}$: The average distances of the lattice points
  change. The parallelogram remains closed.}
\end{figure}

The full description of the corresponding scheme requires the
formalism of fiber bundles and connections on fiber bundles, see,
among many others, the early work on this subject by Cho
\cite{Cho1976}, also the recent important work of Tresguerres
\cite{Tresguerres2007}, and the references given therein.  Here we
only briefly formulate the general ideas and basic results of the
translational gauge approach.

In accordance with the general gauge-theoretic scheme, to the three generators
$P_\alpha$ of the translation group there corresponds a Lie algebra-valued 1-form 
$\Gamma^{(T)} = \Gamma_{i}^{(T)\alpha}P_\alpha\,dx^i$ as the translational 
gauge field potential. Under translations $y^\alpha\rightarrow y^\alpha + 
\epsilon^\alpha$ in the affine tangent space, it transforms like a connection 
\begin{equation}
\delta \Gamma_{i}^{(T)\alpha} = - \partial_i\epsilon^\alpha.\label{GTtr} 
\end{equation}
Since $T_3$ is Abelian, i.e., translations commute with each other,
there is no homogeneous term in this transformation law. Thus, it
resembles the phase transformation of an electromagnetic potential.
For the same reason, the gauge field strength $F^{(T)\alpha} =
d\Gamma^{(T)\alpha} = {\frac 12}\,F_{ij}^{(T)\alpha}\,dx^i\wedge dx^j$
is formally reminiscent of a generalized electromagnetic field strength.
This analogy was extensively used by Itin \cite{Itin0,Itin}.

\begin{figure}\label{dislFig3}
\begin{center}
\includegraphics[width=9cm]{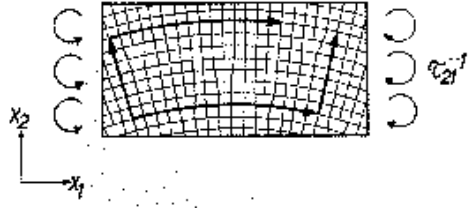}
\end{center}
\caption{Deformation of a cubic crystal by edge dislocations of type
  $\a_{12}{}^1$: The relative orientations of the lattice plains in
  2-direction change. A vector in $x_2$-direction will rotate, if
  parallelly displaced along the $x_1$-direction. As a consequence a
  contortion $\kappa_{112}$ emerges and the closure failure occur of
  the ``infinitesimal'' parallelogram.}
\end{figure}

\begin{figure}\label{RCspace}
\begin{center}
\includegraphics[width=7.8cm]{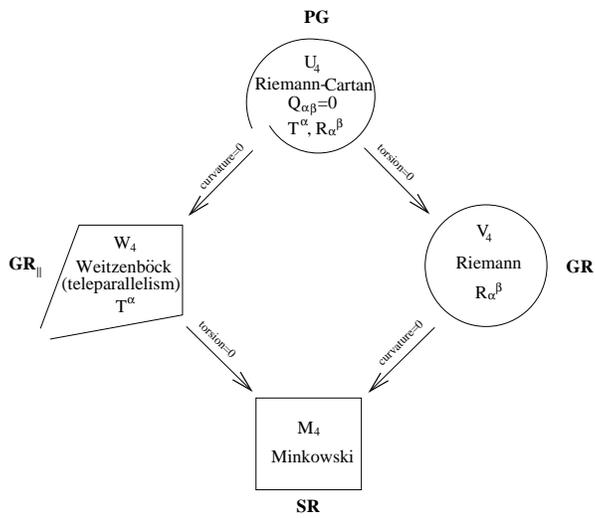}
\caption{ A Riemann-Cartan space and its special cases of a
  Weitzenb\"ock and a Riemannian space.}
\end{center}
\end{figure}
 
In addition to the translational gauge field, another important
structure is a field $\xi^\alpha$ defined as a local section of the
affine tangent bundle.  Geometrically, this field determines the
``origin" of the affine spaces; it is known as Cartan's ``radius
vector''.  Under the gauge transformation (translation) it changes as
$\xi^\alpha \rightarrow \xi^\alpha + \epsilon^\alpha$. However, the
combination $e_{i}{}^{\alpha} = \partial_{i}\xi^{\alpha} +
\Gamma_{i}^{(T)\alpha}$ is obviously {\it gauge invariant}, see
\cite{PRs}, Eq.(3.3.1). In a rigorous gauge-theoretic framework, the
1-form $\vartheta^\alpha = e_{i}{}^{\alpha}
dx^i=d\xi^\a+\Gamma^{(T)\a}$ arises as the nonlinear translational
gauge field with $\xi^\alpha$ interpreted as the Goldstone field
describing the spontaneous breaking of the translational symmetry.

We can consistently treat $\vartheta^\alpha = e_{i}{}^{\alpha} dx^i$
as the coframe of our 3D manifold. Then the translational gauge field
strength is actually the anholonomity 2-form of this coframe:
$F^{(T)\alpha} = d\Gamma^{(T)\alpha} = d\vartheta^\alpha$.
Collecting our results, we have the deformation measures 
\begin{eqnarray}\label{radius}
e_{i}{}^{\alpha}\, &=&\, \partial_{i}\xi^{\alpha} +
\Gamma_{i}^{(T)\alpha}\,,\\
F^{(T)\alpha}\,& = &\, d\Gamma^{(T)\alpha} = d\vartheta^\alpha\,.\label{anh}
\end{eqnarray}

If, in linear approximation, we compare these measures with the
Cosserat deformation measure (\ref{Coss1}),(\ref{Coss2}), then we
find, in generalization of the Cosserat structure,
\begin{eqnarray}\label{compare1}
  e_i{}^\a\,&\rightarrow &\, \b_{ij}\quad\mbox{(distortion)}\,,\\ \label{compare2}
  \xi^\a\,&\rightarrow&\,u_i\quad\mbox{(displacement)}\,,\\ \label{compare3}
  \Gamma_{i}^{(T)\alpha}\,&\rightarrow &\, \omega_{ij}\quad(!)\,,\\
 \label{compare4}
 F_{ij}^{(T)\alpha}\,&\rightarrow &\, \kappa_{kji}\quad\mbox{(contortion)}\,.
\end{eqnarray}
Here $F_{ij}^{(T)\alpha}\sim\alpha_{ijk}$ represents the dislocation
density (torsion). Hence (\ref{compare4}) represents Nye's relation
(\ref{contortion}), and the second deformational measure of
dislocation theory with its 9 independent components corresponds to
the contortion of the Cosserat theory. However, as we can recognize
from (\ref{compare3}), the dislocated continuum requires a more
general description. The 3 component Cosserat rotation $\omega_{ij}=
-\omega_{ji}$ is substituted by the asymmetric 9 component
(translational gauge) potential $ \Gamma_{i}^{(T)\alpha}$. Still, the
distortion $\b_{ij}$ carries also 9 independent components and the
corresponding static response is represented by the asymmetric force
stress $\overline{\Sigma}_{ij}\sim\d{\cal H}/\d \b_{ij}$.

If the second deformation measure in dislocation theory were, similar
to the Cosserat theory, the gradient of $ \Gamma_{i}^{(T)\alpha}$,
i.e., $\partial_j\ \Gamma_{i}^{(T)\alpha}$, it would have 27
independent components and the static responses would be represented
hyperstresses with and without moments, see \cite{GronHehl}. However,
as it turns out, see (\ref{compare4}) --- and this is very decisive
--- it is the dislocation density (torsion), i.e., the {\it curl} of
$\Gamma_{i}^{(T)\alpha}$, with only 9 independent components that
plays a role. For this reason, the static response in dislocation
theory are again, as in a Cosserat continuum, just spin moment
stresses $\overline{\tau}_{ijk}\sim\d{\cal H}/\d K_{kji}$, see
\cite{HK}. Note that $\overline{\tau}_{ijk}$ is equivalent
to\footnote{In 4D it is called the spin energy potential, see
  \cite{PRs}, Eqs.(5.1.24) and (5.1.22).}
$\overline{\mu}_{ijk}\sim\d{\cal H}/\d \a_{kji}$. Thus, in dislocation
theory as well as in the Cosserat continuum, we have the same type of
stresses $\overline{\Sigma}_\a$ and $\overline{\tau}_{\a\b}$ in spite
of the newly emerging 9 component field $\Gamma_{i}^{(T)\alpha}$.

Continuum theories of {\it moving dislocations} are still a developing
subject, see, e.g., Lazar \cite{Lazar} and Lazar \& Anastassiadis
\cite{LazarA} (and the literature quoted therein). Probably it is fair
to say that they didn't find too many real applications so far.
Nevertheless, the identification of the dislocation density with the
torsion is invariably a cornerstone of all these theories.

\section{Translational gauge theory of gravity}

The construction of a translation gauge theory does not depend on the
dimension of the underlying space. Hence we can take in 4D spacetime
the same fundamental formulas (\ref{radius}),(\ref{anh}).
Incidentally, the construction of the gauge theory for the group of
translations is quite nontrivial because the local spacetime
translations look very similar to the diffeomorphisms of spacetime.
They are, however, different
\cite{TresguerresMielke2000,Tresguerres2007}. The underlying
geometrical structure of the theory is, as explained in the previous
section, the affine tangent bundle. The corresponding translational
connection is the 1-form $\Gamma_{i}^{(T)\alpha}dx^i$ with the
transformation law (\ref{GTtr}). Now, however, the Latin and Greek
indices run {}from 0 to 3.

With the help of the Goldstone type field $\xi^\alpha$, the
translational gauge field gives rise to the coframe $\vartheta^\alpha
= e_{i}{}^{\alpha} dx^i$ as described in (\ref{radius}). The
anholonomity 2-form $F^{(T)\alpha}$ is the corresponding translational
gauge field strength (\ref{anh}). The gravitational theories based on
the coframe as the fundamental field have long history. The early
coframe (or so-called vierbein, or tetrad, or teleparallel) gravity
models were developed by M\o{}ller \cite{Moller}, Pellegrini and
Pleba\'nski \cite{Pele}, Kaempfer \cite{Kaempfer}, Hayashi and
Shirafuji \cite{HSh79}, to mention but a few. The first fiber bundle
formulation was provided by Cho \cite{Cho1976}. The dynamical contents
of the model was later studied by Schweizer et al.\ \cite{Schweizer},
Nitsch and Hehl \cite{Nitsch1979}, Meyer \cite{Meyer}, and more recent
advances can be found in Aldrovandi and Pereira
\cite{AldrovandiPereira}, Andrade and Pereira \cite{AndradePereira},
Gronwald \cite{GronwaldIJMPD}, Itin \cite{Itin,Itin2}, Maluf and da
Rocha-Neto \cite{Maluf}, Muench \cite{Muench97}, Obukhov and Pereira
\cite{OP}, and Schucking and Surowitz
\cite{Schuecking,SchueckingSurowitz}.

The Yang-Mills type Lagrangian 4-form for the translational gauge field 
$\vartheta^\alpha$ is constructed as the sum of the quadratic invariants 
of the field strength: 
\begin{equation}
\tilde{V}(\vartheta,d\vartheta) = -\,\frac{1}{2\kappa}\,F^{(T)\alpha}\wedge 
{}^\star\left(\sum_{I=1}^3\,a_I\,{}^{(I)}F^{(T)}_{\alpha}\right). \label{V1}
\end{equation}
Here $\kappa=8\pi G/c^3$, and ${}^\star$ denotes the Hodge dual of the
Minkowski flat metric $g_{\alpha\beta}= o_{\alpha\beta} :={\rm
  diag}(-1,1,1,1)$, that is used also to raise and lower the Greek
(local frame) indices. As it is well known, we can decompose the field
strength $F^{(T)\alpha}$ into the three irreducible pieces of the
field strength:
\begin{eqnarray}\label{Fi1}
{}^{(1)}F^{(T)\alpha}&:=&F^{(T)\alpha}-{}^{(2)}F^{(T)\alpha}
-{}^{(3)}F^{(T)\alpha},\\
{}^{(2)}F^{(T)\alpha}&:=&\frac{1}{3}\,\vartheta^\alpha\wedge\left(e_\beta\rfloor
F^{(T)\beta}\right),\label{Fi2}\\
{}^{(3)}F^{(T)\alpha}&:=&\frac{1}{3}\,e^\alpha\rfloor\left(\vartheta^\beta\wedge
F^{(T)}_\beta\right),\label{Fi3}
\end{eqnarray}
i.e., the tensor part, the trace, and the axial trace, respectively. 

There are three coupling constants in this theory, in general: $a_1, a_2, a_3$.
In accordance with the general Lagrange-Noether scheme \cite{GronwaldIJMPD,PRs} 
one derives {}from (\ref{V1}) the translational excitation 2-form and 
the canonical energy-momentum 3-form:
\begin{eqnarray}
\tilde{H}_{\alpha} = -\,{\frac {\partial \tilde{V}} {\partial F^{(T)\alpha}}} &=&
\,{1\over \kappa}\,{}^\star\!\left(\sum_{I=1}^3\,a_I\,{}^{(I)}F^{(T)}_{\alpha}
\right), \label{defH} \\
\tilde{E}_\alpha = {\frac {\partial \tilde{V}} {\partial \vartheta^\alpha}}
&=& e_\alpha\rfloor \tilde{V} + (e_\alpha\rfloor F^{(T)\beta})\wedge 
\tilde{H}_\beta. \label{defE}
\end{eqnarray}
Accordingly, the variation of the total Lagrangian $L=\tilde{V} + 
L_{\rm mat}$ with respect to the tetrad results in the gravitational 
field equations
\begin{equation}
d\tilde{H}_\alpha-\tilde{E}_\alpha=\Sigma_\alpha, \label{fe1}
\end{equation}
with the canonical energy-momentum current 3-form of matter 
\begin{equation}
\Sigma_\alpha:=\frac{\delta L_{\rm mat}}{\delta \vartheta^\alpha}
\label{defsigma}
\end{equation}
as the source. 

The coframe models do not possess any other symmetry except the diffeomorphism
invariance and the invariance under the {\it rigid} Lorentz rotations of the
tetrads. However, for a special choice of the coupling constants,
\begin{equation}
a_1 = 1,\qquad a_2 = -\,2,\qquad a_3 = -\,{\frac 12},\label{cTE}
\end{equation}
the field equations turn out to be invariant under the {\it local Lorentz} 
transformations $\vartheta^\alpha\longrightarrow L^\alpha{}_\beta(x)
\vartheta^\beta$ with the matrices $L^\alpha{}_\beta(x)$ arbitrary functions
of the spacetime coordinates. At the same time, one can demonstrate that 
the tetrad field equations (\ref{fe1}) are then recast {\it identically} into 
the form of Einstein's equation
\begin{equation}
\frac{1}{2\kappa}\,\eta_{\alpha\beta\gamma}\wedge \tilde{R}^{\beta\gamma}
= \Sigma_\alpha.\label{GReinstein}
\end{equation}
Here $\tilde{R}_\alpha{}^\beta = d\tilde{\Gamma}_\alpha{}^\beta + 
\tilde{\Gamma}_\gamma{}^\beta\wedge\tilde{\Gamma}_\alpha{}^\gamma$ is the 
Riemannian curvature of the Christoffel connection 
\begin{equation}
\tilde{\Gamma}_{\alpha\beta}:=\frac{1}{2}\left[e_\alpha\rfloor
F^{(T)}_\beta-e_\beta\rfloor F^{(T)}_\alpha-(e_\alpha\rfloor e_\beta\rfloor 
F^{(T)}_\gamma)\wedge \vartheta^\gamma\right].\label{deftildeGamma}
\end{equation}
For that reason, the coframe gravity model with the choice (\ref{cTE})
is usually called a {\it teleparallel equivalent} of general
relativity theory.

\section{Einstein-Cartan theory of gravity}\label{ECT}

Einstein-Cartan (EC) theory is an extension of Einstein's general relativity,
in which the local Lorentz symmetry, which appears to be accidental in the
teleparallel equivalent model above, is taken seriously as a fundamental 
feature of the gravitational theory. 

One can naturally arrive to the EC-theory using the heuristic 
arguments based on the mapping of the Noether to Bianchi identities, as shown 
in McCrea et al.\ \cite{HMcCrea,McMap}. Similar are the thoughts of Ruggiero 
and Tartaglia \cite{RT}, who consider the EC-theory as a defect type theory;
see also Hammond \cite{Hammond}, Ryder and Shapiro \cite{Lewis}, and
Trautman \cite{TrautmanHeld,Trautman}.

However, the most rigorous derivation is based on the gauge approach for the
Poincar\'e group (see \cite{GronwaldIJMPD,PRs}, for example), in which the 
gauge potentials are the coframe $\vartheta^\alpha$ and the Lorentz connection
$\Gamma_\alpha{}^\beta$. They correspond to the translational and the Lorentz
subgroups of the Poincar\'e group, respectively. 

The dynamics of the gravitational field is described in this model by the
Hilbert-Einstein Lagrangian plus, in general, a cosmological term:
\begin{equation}\label{HEV}
  V = - \,{\frac 1{2\kappa}}\left(\eta_{\alpha\beta}\wedge  R^{\alpha\beta}
    - 2\lambda\eta\right).
\end{equation} 
The field equations arise {}from the variations of the total
Lagrangian $V_{\rm tot} = V + L_{\rm mat}$ with respect to the coframe
and connection, see Sciama \cite{Sciama} and Kibble \cite{Kibble}:
\begin{eqnarray}
{\frac {1}{2}}\,\eta_{\alpha\beta\gamma} \wedge R^{\beta\gamma}
- \lambda\eta_\alpha\, &=&\, \kappa\Sigma_\alpha,\label{einstein1}\\
{\frac {1}{2}}\,\eta_{\alpha\beta\gamma}\wedge T^{\gamma}\, &=&\, 
\kappa\tau_{\alpha\beta}.\label{cartan1}
\end{eqnarray}
Here in addition to the canonical energy-momentum current (\ref{defsigma}),
the canonical spin current 3-form of matter 
\begin{equation}
\tau^\alpha{}_\beta :=\frac{\delta L_{\rm mat}}{\delta \Gamma_\alpha{}^\beta}
\label{deftau}
\end{equation}
arises as the source of the gravitational field. Two sources $\Sigma_\alpha$
and $\tau^\alpha{}_\beta$ satisfy the identities (``covariant conservation
laws") that follows {}from the Noether theorem for the invariance of the theory
under diffeomorphisms and the local Lorentz group:
\begin{eqnarray}\label{NoeT}
D\Sigma_\alpha\, &=&\, (e_\alpha\rfloor T^\beta)\wedge\Sigma_\beta+
(e_\alpha\rfloor R_{\beta\g})\wedge\tau^{\beta\g}\,,\\
D\tau_{\alpha\beta} &+& \vartheta_{[\alpha}\wedge\Sigma_{\beta]} =0.\label{NoeL}
\end{eqnarray}

When the matter has no spin, $\tau^\alpha{}_\beta = 0$, the second
(Cartan's) field equation (\ref{cartan1}) yields the zero spacetime
torsion, $T^\alpha =0$.  As a result, the Riemann-Cartan curvature
$R^{\beta\gamma}$ reduces to the Riemannian curvature
$\tilde{R}^{\beta\gamma}$, and the first field equation
(\ref{einstein1}) reduces to Einstein's equation (\ref{GReinstein}) of
general relativity theory. Physical effects of classical and quantum
matter in the EC-theory are overviewed in \cite{RMP,Shapiro}. There
emerges, as compared to general relativity, an additional spin-spin
contact interaction of gravitational origin that only plays a role at
extremely high matter densities.

Blagojevi\'c et al.\ \cite{MilutinTorsion} found in 3D gravity with
torsion an interesting quantum effect: The black hole entropy depends
on the torsional degrees of freedom.

\section{Poincar\'e gauge theory and metric-affine gravity}

Einstein-Cartan theory, outlined in Sec.~\ref{ECT}, represents a degenerate
Poincar\'e gauge model in which spin couples algebraically to the Lorentz
connection. As a result, torsion is a nonpropagating field and vanishes
identically outside the material sources. 

Things are however different in the Yang-Mills type models of the
Poincar\'e gravity based on the quadratic Lagrangians in torsion and
curvature. These models are discussed by Hehl \cite{College},
Ponomariov and Obukhov \cite{Pono}, Gronwald and Hehl \cite{Erice},
see also a recent review by Obukhov \cite{Yuridiffgeo}.

The general Lagrangian which is at most {\it quadratic} (q) in the
Poincar\'e gauge field strengths -- in the torsion and the curvature
-- reads
\begin{eqnarray} 
V_{\rm q}&=&
-\,\frac{1}{2\kappa}\,\left[a_0\,R^{\alpha\beta}\wedge\eta_{\alpha\beta}
-2\lambda\,\eta + T^\alpha\wedge{}^*\!\left(\sum_{I=1}^{3}a_{I}\,^{(I)}
T_\alpha\right)\right]\nonumber\\
&&-\,\frac{1}{2}\,R^{\alpha\beta}\wedge{}^*\!\left(\sum_{J=1}^{6}b_{J}\,
{}^{(J)}R_{\alpha\beta}\right).\label{lagr}
\end{eqnarray}
We use the unit system in which the dimension of the gravitational constant 
is $[\kappa] = \ell^2$ with the unit length $\ell$. The coupling constants
$a_0,a_1,a_2,a_3$ and $b_1,...,b_6$ are {\it dimensionless}, whereas 
$[\lambda] = \ell^{-2}$. These coupling constants determine the particle
contents of the quadratic Poincar\'e gauge models. The three irreducible parts
of the torsion ${}^{(I)}T_\alpha$ are defined along the pattern 
(\ref{Fi1})-(\ref{Fi3}), whereas the irreducible decomposition of the curvature 
into the six pieces ${}^{(J)}R_{\alpha\beta}$ is given in \cite{PRs}.
The Lagrangian (\ref{lagr}) has the general structure similar to that of the 
Yang--Mills Lagrangian for the gauge theory of internal symmetry group.

The Poincar\'e gauge field equations are derived {}from the total Lagrangian 
$V_{\rm q} + L_{\rm mat}$ {}from the variations with respect to the coframe
and connection. They read explicitly 
\begin{eqnarray} 
DH_{\alpha}- E_{\alpha}\,&=&\,\Sigma_{\alpha}\,,\label{first}\\ 
DH^{\alpha}{}_{\beta}- E^{\alpha}{}_{\beta}\,&=&\,\tau^{\alpha}{}_{\beta}\,.
\label{second}
\end{eqnarray} 
The right-hand sides describe the material sources of the Poincar\'e gauge
gravity: the canonical energy--momentum (\ref{defsigma}) and the spin 
(\ref{deftau}) three--forms. The left-hand sides are constructed {}from the
{\it gauge field momenta} 2-forms
\begin{equation} 
H_{\alpha} := - {\frac{\partial V_{\rm q}}{\partial T^{\alpha}}}\,,\qquad  
H^{\alpha}{}_{\beta} := 
- {\frac{\partial V_{\rm q}}{\partial R_{\alpha}{}^{\beta}}}\,,\label{HH}
\end{equation}  
and the {\it canonical} $3$--forms of the energy-momentum and spin of the
gauge gravitational field 
\begin{eqnarray}\label{Ea0}
E_{\alpha}\, &:=&\, {\frac{\partial V_{\rm q}}{\partial\vartheta^{\alpha}}} = 
e_{\alpha}\rfloor V_{\rm q} + (e_{\alpha}\rfloor T^{\beta})\wedge H_{\beta} 
+ (e_{\alpha}\rfloor R_{\beta}{}^{\gamma})\wedge H^{\beta}{}_{\gamma},\\
E^{\alpha\beta}\,&:=&\, {\frac{\partial V_{\rm q}}{\partial\Gamma_{\alpha\beta}}}
= - \vartheta^{[\alpha}\wedge H^{\beta]}. \label{Eab}
\end{eqnarray}

The class of gravitational models (\ref{lagr}) has a rich geometrical and
physical structure. Depending on the choice of the coupling constants 
$a_0,a_1,a_2,a_3$ and $b_1,...,b_6$, the field equations (\ref{first})
and (\ref{second}) admit black hole, cosmological and wave solutions that
generalize the general-relativistic solutions of Einstein's theory at small 
distances. On the large time and space scales, the physical predictions of
the Poincar\'e gravity generally agree with results of the general relativity,
see \cite{College,Erice,PRs,Yuridiffgeo}.

The Cosserat medium in elasticity theory and the physical sources in
the Poincar\'e gauge gravity deal with the material continua and
bodies, the elements of which have rigid microstructure. A further
generalization is possible when the matter elements possess {\it
  deformable microstructure}. In elasticity theory this is the case,
for example, in Mindlin's 3-dimensional continuum with microstructure
\cite{Mindlin}. In 4 dimensions, the corresponding counterpart arises
as {\it metric-affine gravity} (MAG) theory. The proper framework is
then the gauge theory based on the general affine symmetry group
\cite{PRs}. The geometry of such an elastic medium and of the
spacetime in MAG is characterized, in addition to the curvature and
torsion, by a nontrivial {\it nonmetricity}.

\section{Cartan's spiral staircase: A 3D Euclidean model for a space
  with torsion}

\begin{figure}\label{staircase}
\begin{center}
\includegraphics[width=6.5cm]{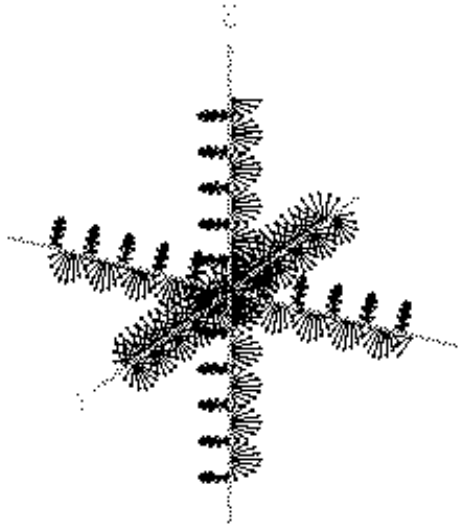}
\caption{ {\em Cartan's spiral staircase,} see Garc{\'\i}a et al.\
    \cite{Garcia}. Cartan's rules \cite{Cartan1922} for the
  introduction of a non-Euclidean connection in a 3D Euclidean space
  are as follows: (i) A vector which is parallelly transported along
  itself does not change (cf.\ a vector directed and transported in
  $x$-direction). (ii) A vector that is orthogonal to the direction of
  transport rotates with a prescribed constant `velocity'' (cf.\ a
  vector in $y$--direction transported in $x$--direction).  The
  winding sense around the three coordinate axes is always positive.}
\end{center}
\end{figure}

Apparently in order to visualize torsion in a simple 3D model, see
Fig.13, Cartan proposed a certain construction that, in his own
(translated) words of 1922 \cite{Cartan1922}, reads as follows:
\begin{quote}
  ``\dots imagine a space F which corresponds point by point with a
  Euclidean space E, the correspondence preserving distances. The
  difference between the two space is following: two orthogonal triads
  issuing {}from two points A and A' infinitesimally nearby in F will be
  parallel when the corresponding triads in E may be deduced one {}from
  the other by a given helicoidal displacement (of right--handed sense,
  for example), having as its axis the line joining the origins. The
  straight lines in F thus correspond to the straight lines in E: They
  are geodesics. The space F thus defined admits a six parameter group
  of transformations; it would be our ordinary space as viewed by
  observers whose perceptions have been twisted.  Mechanically, it
  corresponds to a medium having constant pressure and constant internal
  torque.''
\end{quote}

One can show \cite{Garcia} that Cartan's prescription yields a trivial
coframe and a constant connection,
\begin{equation}\label{stair1}
  \vartheta^\alpha=\delta_i^\alpha\,dx^i\,,\qquad
  \Gamma^{\alpha\beta}=\frac{\cal T}{\ell}\,\eta^{\alpha\beta}\,,
\end{equation}
with the 1-form $\eta^{\a\b}={}^\star\left(\vt^\a\wedge\vt^\b \right)$
and $^\star$ as the Hodge star operator; moreover, $\cal T$ and $\ell$
are constants. The components of the connection are totally
antisymmetric,
$\Gamma_{\gamma\alpha\beta}=e_\gamma\rfloor\Gamma_{\alpha\beta}=({\cal
  T}/\ell)\,\eta_{\gamma\alpha\beta}$. Thus, autoparallels and
geodesics coincide. Accordingly, in the spiral staircase, extremals
are {\em Euclidean} straight lines. This is apparent in Cartan's
construction. By simple algebra we find for the torsion, the
Riemannian curvature, and the Riemann-Cartan curvature, respectively,
\begin{equation}\label{stair2}
  T^{\alpha} = 2\,\,\frac{{{\cal T}}}{\ell} \,^\star \vt^{\alpha}\,,
  \qquad\widetilde{R}^{\alpha\beta} =0\,,\qquad R^{\alpha\beta} =-
  \frac{{\cal T}^2}{\ell^2} \, \vartheta^{\alpha}\wedge \vartheta^{\beta}\,.
\end{equation}

For a {\it solid state physicist} it is immediately clear that the
geometry in Fig.13 represents a set of three perpendicular constant
`forests' of {\it screw} dislocations of equal strength. Hence Cartan
thought in terms of screw dislocations without knowing them! Of
course, the totally antisymmetric part of the dislocation density
$\a_{[ijk]}$ is an irreducible piece of the torsion which has one
independent component. Wouldn't it be interesting to find this spiral
staircase as an exact solution in dislocation gauge theory? Since only
one irreducible piece of the dislocation density (torsion) is
involved, this should be possible.

Cartan apparently had in mind a 3D space with Euclidean signature.
For an alternative interpretation of Cartan's spiral staircase we
consider the 3D Einstein--Cartan field equations without cosmological
constant:
\begin{eqnarray}
  \frac{1}{2} \, \eta_{\alpha\beta\gamma} \, R^{\beta\gamma}
\,&=&\,\ell \, \overline{\Sigma}_\alpha \,,\\ \frac{1}{2} \, \eta_{
    \alpha\beta\gamma} \, T^\gamma\, &=&\, \ell \, \overline{\tau}_{\alpha\beta} \,.
\end{eqnarray}
The coframe and the connection of (\ref{stair1}), Euclidean signature
assumed, form a solution of the Einstein--Cartan field equations {\em
  with matter} provided the energy--momentum current (for Euclidean
signature the force stress tensor ${\mathfrak t}_\alpha{}^\beta$) and
the spin current (here the torque or spin moment stress tensor
${\mathfrak s}_{\alpha\beta}{}^\gamma$) are constant,
\begin{equation}
  \overline{\Sigma}_{\alpha} =:{\mathfrak t}_\alpha{}^\beta\,\eta_\beta = -
  \frac{{\cal T}^2}{\ell^3}
\,\eta_\alpha\,\quad{\rm and}\quad
 \overline{\tau}_{\alpha\beta} =:{\mathfrak s}_{\alpha\beta}{}^\gamma\,\eta_\gamma =
 - \frac{\cal T}{\ell^2} \,
  \vartheta_{\alpha\beta}\,.
\end{equation}
Inversion yields
\begin{equation}
  {\mathfrak t}_\alpha{}^\beta= -\frac{{\cal
      T}^2}{\ell^3} \, 
\delta_\alpha^\beta\,,\qquad {\mathfrak
    s}_{\alpha\beta\gamma}= -\frac{\cal T}{\ell^2} \,\eta_{\alpha \beta
    \gamma}\,.
\end{equation}
We find a constant hydrostatic {\em pressure} $ -{\cal T}^2/\ell^3$
and a constant {\em torque} $-{\cal T}/\ell^2$, exactly as foreseen by
Cartan. 

By studying the spiral staircase and reading also more in the Cartan
book \cite{Cartan1986}, it becomes clear that Cartan's intuition
worked in 3D (and not in 4D). This led Cartan to a decisive mistake
in this connection. Take the energy-momentum law in a 4D RC-space, if
the mater field equation is fulfilled, see (\ref{NoeT}),
\begin{equation}\label{Noe4}
D\Sigma_\alpha
  =(e_\alpha\rfloor T^\beta)\wedge\Sigma_\beta+
(e_\alpha\rfloor R_{\beta\g})\wedge\tau^{\beta\g}\,,\qquad\mathbf{4D}\,.
\end{equation}
Note the Lorentz type forces on the right-hand-side, in particular the
last term representing a Mathisson-Papapetrou type of force with
curvature $\times$ spin. However, straightforward algebra yields, for
3D,
\begin{equation}\label{Noe3}
  D\overline{\Sigma}_\alpha=0\,,
  \qquad\mathbf{3D}\,.
\end{equation}
Cartan assumed incorrectly that (\ref{Noe3}) is also valid in four
dimensions. For that reason he ran into difficulties with his 4D
gravitational theory that includes (\ref{Noe3}) and came, after his
1923/1924 papers, never back to his (truncated
Einstein-Cartan-)theory.  Hence intuition (without algebra) can even
lead the greatest mathematical minds astray.

\section{Some controversial points}

In more physically oriented papers, the authors are often open minded
and treat the question of the possible existence of a torsion as a
dynamical one. Hanson and Regge \cite{HansonRegge}, e.g., open their
paper with the statement: {\it ``We suggest that the absence of torsion
  in conventional gravity could in fact be dynamical. A gravitational
  Meissner effect might produce instanton-like vortices of nonzero
  torsion concentrated at four-dimensional points...''} Accordingly
they study certain dynamical models in order to find a possible
answer for this question. We don't follow this train of
thought. However, such a model building is a desirable feature.
 
In contrast, in the literature there are numerous statements about a
possible torsion of the spacetime manifold that don't stand a closer
examination. Let us quote some examples:

\begin{enumerate}

\item {\it Ohanian and Ruffini} \cite{Ohanian} claim that the
  Einstein-Cartan theory is defective, see ref.\cite{Ohanian}, pp.\
  311 and 312. Since this is a widely read and, otherwise, excellent
  textbook, we would just like to comment on their arguments, see also
  \cite{Hehl1997}:

\emph{``If $\,\Gamma^\beta{}_{\nu\mu}$ were not symmetric, the parallelogram
would fail to close.  This would mean that the geometry of the curved
spacetime differs {}from a flat geometry even on a small scale -- the
curved spacetime would not be approximated locally by a flat
spacetime."}

Equation (\ref{propo}), see also the paper of Hartley \cite{Hartley},
disproves the Ohanian and Ruffini statement right away. In
(\ref{propo}) it is clearly displayed that the Riemann-Cartan geometry
is Euclidean `in the infinitesimal'. And this was, as we discussed in
Sec.2, one of the guiding principles of Cartan.

\emph{``...we do not know the `genuine' spin content of elementary
    particles...''}

  According to present day wisdom, matter is built up {}from quarks
  and leptons. No substructures have been found so far. According to
  the mass-spin classification of the Poincar\'e group and the
  experimental information of lepton and hadron collisions etc.,
  leptons and quarks turn out to be fermions with spin 1/2 (obeying
  the Pauli principle). As long as we accept the (local) Poincar\'e
  group as a decisive structure for describing elementary particles,
  there can be no doubt what spin really is. And abandoning the
  Poincar\'e group would result in an overhaul of (locally valid)
  special relativity theory.

  The nucleon is a composite particle and things related to the
  build-up of its spin are not clear so far. But we do know that we
  can treat it as a fermion with spin $1/2$. As long as this can be
  taken for granted, at least in an effective sense, we know its spin
  and therefore its torsion content.

\item {\it Carroll} \cite{Carroll} argues in his book on p.190 as follows:

  \emph{``...Thus, we do not really lose any generality by considering
    theories of torsion-free connections (which lead to GR) plus any
    number of tensor fields, which we can name what we like. Similar
    considerations...''}

  (i) This opinion is often expressed by particle physicists who don't
  think too profoundly about geometry. As we saw in Secs.1 and 2,
  the torsion tensor is not {\it any} tensor, but it is a particular
  tensor related to the {\it translation group.} A torsion tensor
  cracks infinitesimal parallelograms, see Fig.1. A parallelogram is
  deeply related to the geometry of a manifold with a linear
  connection. The closure failure of a parallelogram can only be
  created by a distinctive geometrical quantity, namely the torsion
  tensor --- and not by any other tensor.  This fact alone makes
  Carroll's argument defective.

  (ii) Another way of saying this is that torsion affects the Bianchi
  identities (\ref{Bianchi}). This cannot be done by any other tensor,
  apart from the curvature tensor. Moreover, as we saw in Sec.5, the
  torsion is the field strength belonging to the translation group.

  (iii) As in particular Sciama \cite{Sciama} has shown, an independent
  Lorentz connection couples to the spin of a matter field in a
  similar way as the coframe couples to the energy-momentum of matter.
  This shows too that a splitting off of the Levi-Civita connection is
  of no use in such a context. The Einstein-Cartan theory of gravity
  is a viable gravitational theory. If one studies its variational
  principle etc., then one will recognize that the splitting technique
  advised by Carroll messes up the whole structure.

  (iv) If one minimally couple to a connection, it is decisive which
  connection one really has. Of course, one can couple minimally to
  the Levi-Civita connection and add later nonminimal $\sim$
  (torsion)$^2$ pieces thereby transforming a minimal to a nonminimal
coupling; also here one messes up the structure. Minimal coupling
would lose its heuristic power.

\item {\it Kleinert and Shabanov} \cite{Kleinert} postulate that a scalar
  particle moves in a Riemann-Cartan space along an autoparallel.
  However, the equations of motion cannot be postulated freely, they
  have rather to be determined from the energy-momentum and the
  angular momentum laws of the underlying theory. Then it turns out
  that a scalar particle can only `feel' the Riemannian metric of
  spacetime, it is totally insensitive to a possibly existent torsion
  (and nonmetricity) of spacetime. This has been proven, e.g., by
  Yasskin and Stoeger \cite{YS}, Ne'eman and Hehl \cite{NH}, and by
  Puetzfeld and Obukhov \cite{PO}.

\item {\it Weinberg} \cite{WeinbergEinstein} wrote an article about
  ``Einstein's mistakes''. In a response, Becker \cite{Becker} argued
  that for ``generalizing general relativity'' one should allow
  torsion and teleparallelism. Weinberg's response \cite{WeinbergPT}
  was as follows:

  \emph{``I may be missing the point of Robert Becker's remarks, but I
    have never understood what is so important physically about the
    possibility of torsion in differential geometry. The difference
    between an affine connection with torsion and the usual
    torsion-free Christoffel symbol is just a tensor, and of course
    general relativity in itself does not constrain the tensors that
    might be added to any dynamical theory.  What difference does it
    make whether one says that a theory has torsion, or that the
    affine connection is the Christoffel symbol but happens to be
    accompanied in the equations of the theory by a certain tensor?
    The first alternative may offer the opportunity of a different
    geometrical interpretation of the theory, but it is still the same
    theory.''}

  This statement of Weinberg was answered by one of us, see
  \cite{WeinbergPT}. We argued, as in this essay, that torsion is
  related to the translation group and that it is, in fact, the
  translation gauge field strength. Moreover, we pointed out the
  existence of a new spin-spin contact interaction in the EC-theory
  and that torsion could be measured by the precession of nuclear
  spins.

Weinberg's answer was:

 \emph{``Sorry, I
    still don't get it. Is there any physical principle, such as a
    principle of invariance, that would require the Christoffel symbol
    to be accompanied by some specific additional tensor? Or that
    would forbid it? And if there is such a principle, does it have
    any other testable consequences?''}

  The physical principle Weinberg is looking for is {\it translational
    gauge invariance}, see Sec.6. And the testable consequences are
  related to the new spin-spin contact interaction and to the
  precession of elementary particle spins in torsion fields.

\item {\it Mao, Tegmark, Guth, and Cabi} \cite{Mao} claim that torsion can
  be measured by means of the Gravity Probe B experiment. This is
  totally incorrect since the sensitive pieces of this gyroscope
  experiment, the rotating quartz balls, don't carry uncompensated
  elementary particle spin. If the balls were made of polarized
  elementary particle spins, that is, if one had a {\it nuclear
    gyroscope}, see Simpson \cite{Simpson}, as they were constructed
  for inertial platforms, then the gyroscope would be sensitive to
  torsion. As mentioned in the last point regarding Kleinert et al.,
  an equation of motion in a general relativistic type of field theory
  has to be derived from the energy-momentum and angular momentum
  laws, see Yasskin and Stoeger \cite{YS} and Puetzfeld and Obukhov
  \cite{POPRL,PO}. Then it turns out that measuring torsion requires
  elementary spin --- there is no other way.

\item {\it Torsion in string theory?} Quite some time ago it was
  noticed by Scherk and Schwarz \cite{ScherkSchwarz} that the
  low-energy effective string theory can be elegantly reformulated in
  geometrical terms by using a non-Riemannian connection. The graviton
  field, the dilaton field, and the antisymmetric tensor field (2-form
  $B$), which represent the massless modes of the closed string, then
  give rise to a spacetime with torsion and nonmetricity. In
  particular, the 3-form $H=dB$ is interpreted in this picture as one
  of the irreducible parts (namely, the axial trace part, cf.\
  (\ref{Fi3})) of the spacetime torsion. Later this idea was extended
  to interpret the dilaton field as the potential for the (Weyl)
  nonmetricity, see \cite{Dereli,Saa,Sazdovic}, for example.

  Another formal observation reveals certain mathematical advantages
  in discussing compactification schemes with torsion for the
  higher-dimensional string models, see \cite{Braaten,Pol2}.

  It is however unclear whether some fundamental principle or model
  underlies these formal observations. The geometrical interpretation
  of this kind is certainly interesting, but one should take it with a
  grain of salt. The qualitative difference (from the elastic models
  with defects and the gauge gravity models) is in the fact that the
  field $H$, although viewed as torsion, is not an independent
  variable in this approach, but it arises from the potential 2-form
  $B$. Consistent with this view is Polchinski's definition of string
  torsion in his glossary, see \cite{Pol2}, p.514: ``{\bf torsion}
  {\it a term applied in various 3-form field strengths, so called
    because they appear in covariant derivatives in combination with
    the Christoffel connection.}''

  Thus, the notion of torsion in string theory is used in an
  unorthodox way and should not be mixed up with Cartan's torsion of
  1922.

\item In the past, there have been several attempts to relate the
  torsion of spacetime to electromagnetism. A recent approach is the
  one of Evans \cite{Evans2003,Evans2005a}, who tried to construct a
  unified field theory. As we have seen in Secs.1 and 2, torsion is
  irresolvably tied to the notion of a translation. Thus, torsion has
  nothing to do with internal (unitary) symmetry groups. We have shown
  in two separate papers \cite{AssessI,AssessII} that Evans' theory is
  untenable.

\end{enumerate}

\section{Outlook}

In three dimensions in dislocated crystals, the equality of the
dislocation density and torsion is an established fact. In four
dimensions, with respect to the experimental predictions, the
Einstein-Cartan theory is a viable gravity model that is presently
indistinguishable from Einstein's general relativity. The contact
character of the spin-connection interaction and the smallness of
Newton's gravitational coupling constant underlies this fact for
macroscopic distances and large times.

Sciama, who was the first, in the year of 1961, to derive the field
equations (\ref{einstein1}),(\ref{cartan1}) in tensor notation
\cite{Sciama}, judged the Einstein-Cartan theory from the point of
view of 1979 as follows (private communication): ``The idea that spin
gives rise to torsion should not be regarded as an ad hoc modification
of general relativity. On the contrary, it has a deep group
theoretical and geometric basis. If the history had been reversed and
the spin of the electron discovered before 1915, I have little doubts
that Einstein would have wanted to include torsion in his original
formulation of general relativity. On the other hand, the numerical
differences which arise are normally very small, so that the
advantages of including torsion are entirely theoretical.''

However, the quadratic Poincar\'e gauge models and their generalizations in
the framework of MAG predict propagating torsion (and nonmetricity) modes 
which can potentially be detected one the extremely small scales (high 
energies). The appropriate physical conditions may occur during the early
stages of the cosmological evolution of the universe, see, e.g., Minkevich
\cite{Minkevich}, Puetzfeld \cite{Puetzfeld}, and Brechet, Hobson, and
Lasenby \cite{Lasenby}.

\subsection*{Acknowledgments}

We would like to thank Valeri Dvoeglazov for inviting us to contribute
to the torsion issue organized by him. One of us (FWH) is very
grateful to Markus Lazar (Darmstadt) for numerous discussions on his
translational gauge theory of dislocations and for providing
appropriate literature on this subject. We thank Dirk Puetzfeld (Oslo)
for comments and for sending us some references.  Financial support
from the DFG (HE 528/21-1) is gratefully acknowledged.\bigskip

\section*{Appendix: Derivation of the deformation measures of a Cosserat
  continuum}

Let us consider a 3D Euclidean space. Its geometrical structure is
determined by the 1-form fields of the coframe ${\stackrel \circ
  \vartheta}{}^\alpha$ and the connection ${\stackrel \circ
  \Gamma}{}_\alpha{}^\beta$.  They satisfy the trivial Cartan
relations:
\begin{eqnarray}\label{zeroT}
d{}{\stackrel \circ \vartheta}{}^\alpha + {\stackrel \circ \Gamma}{}_\beta{}^\alpha
\wedge {\stackrel \circ \vartheta}{}^\beta &=&\,  {\stackrel \circ T}{}^\alpha = 0,\\
d{}{\stackrel \circ \Gamma}{}_\alpha{}^\beta + {\stackrel \circ \Gamma}{}_\gamma
{}^\beta \wedge {\stackrel \circ \Gamma}{}_\alpha{}^\gamma &=&\, 
{\stackrel \circ R}{}_\alpha{}^\beta = 0.\label{zeroR}
\end{eqnarray}
The right-hand sides, given by the torsion and the curvature 2-forms, respectively,
vanish for the Euclidean space.

We now consider an infinitesimal deformation of this manifold produced by the 
``generalized gauge transformation" which is defined as a combination of the 
diffeomorphism and of the local rotation. The diffeomorphism is generated by 
some vector field, whereas the rotation is given by the $3\times 3$ matrix
which acts on the anholonomic (Greek indices) components. We assume that a
deformation is small which means that we only need to consider the infinitesimal
diffeomorphism and rotational transformations. By definition, the
deformation is the sum of the two infinitesimal gauge transformations: 
\begin{eqnarray}
\beta^\alpha &:=& \Delta{\stackrel \circ \vartheta}{}^\alpha = 
\delta_{\rm diff}{\stackrel \circ \vartheta}{}^\alpha + 
\delta_{\rm rot}{\stackrel \circ \vartheta}{}^\alpha,\label{defvta}\\
\kappa_\alpha{}^\beta &:=& \Delta {\stackrel \circ \Gamma}{}_\alpha{}^\beta
= \delta_{\rm diff}{\stackrel \circ \Gamma}{}_\alpha{}^\beta +
\delta_{\rm rot}{\stackrel \circ \Gamma}{}_\alpha{}^\beta\label{defgam}
\end{eqnarray}
Let $u$ be an arbitrary vector field, and we recall that a
diffeomorphism, generated by it, is described by the Lie derivative
along this vector field, i.e., $\delta_{\rm diff} = \ell_u = du\rfloor
+ u\rfloor d$. As for the local rotations, they are given by the
standard transformation formulas,
\begin{equation}\label{rot}
\delta_{\rm rot}{\stackrel \circ \vartheta}{}^\alpha = \varepsilon^\alpha{}_\beta
\,{\stackrel \circ \vartheta}{}^\beta,\quad \delta_{\rm rot} {\stackrel \circ 
\Gamma}{}_\alpha{}^\beta = - {\stackrel \circ D}\varepsilon^\beta{}_\alpha.
\end{equation}
Here ${\stackrel \circ D}$ is the covariant derivative defined by the 
connection ${\stackrel \circ \Gamma}$. 
For the Lie derivative of the coframe we find (with $u^\alpha = u\rfloor
{\stackrel \circ \vartheta}{}^\alpha$)
\begin{eqnarray}
\ell_u{\stackrel \circ \vartheta}{}^\alpha &=& d\,u^\alpha + u\rfloor 
d\, {\stackrel \circ \vartheta}{}^\alpha \nonumber\\
&=& d\,u^\alpha - u\rfloor({\stackrel \circ \Gamma}{}_\beta{}^\alpha \wedge 
{\stackrel \circ \vartheta}{}^\beta) + u\rfloor {\stackrel \circ T}{}^\alpha
\nonumber\\
&=&  d\,u^\alpha + {\stackrel \circ \Gamma}{}_\beta{}^\alpha\,u^\beta - 
(u\rfloor {\stackrel \circ \Gamma}{}_\beta{}^\alpha)\,{\stackrel \circ 
\vartheta}{}^\beta. \label{ellvta}
\end{eqnarray}
We used here (\ref{zeroT}) because the space is Euclidean. Substituting 
(\ref{ellvta}) together with (\ref{rot}) into (\ref{defvta}), we find for 
the translational deformation 
\begin{equation}
\beta^\alpha = {\stackrel \circ D}u^\alpha - \omega^\alpha{}_\beta\,
{\stackrel \circ \vartheta}{}^\beta. \label{trdef}
\end{equation}
Here we introduced $\omega^\alpha{}_\beta := u\rfloor {\stackrel \circ \Gamma}
{}_\beta{}^\alpha - \varepsilon^\alpha{}_\beta$. 

Analogously we have for the Lie derivative of the connection
\begin{eqnarray}
\ell_u {\stackrel \circ \Gamma}{}_\beta{}^\alpha &=& d\,(u\rfloor {\stackrel 
\circ \Gamma}{}_\beta{}^\alpha) + u\rfloor d\,{\stackrel \circ \Gamma}
{}_\beta{}^\alpha \nonumber\\
&=& d\,(u\rfloor {\stackrel \circ \Gamma}{}_\beta{}^\alpha) - u\rfloor
({\stackrel \circ \Gamma}{}_\gamma {}^\alpha \wedge {\stackrel \circ \Gamma}
{}_\beta{}^\gamma ) + u\rfloor {\stackrel \circ R}{}_\beta{}^\alpha\nonumber\\
&=& d\,(u\rfloor {\stackrel \circ \Gamma}{}_\beta{}^\alpha) + {\stackrel \circ 
\Gamma}{}_\gamma{}^\alpha (u\rfloor {\stackrel \circ \Gamma}{}_\beta{}^\gamma)
- {\stackrel \circ \Gamma}{}_\beta{}^\gamma (u\rfloor {\stackrel \circ \Gamma}
{}_\gamma{}^\alpha).\label{ellgam}
\end{eqnarray}
We again used here (\ref{zeroR}) for the Euclidean space. Now, substituting
(\ref{ellgam}) together with (\ref{rot}) into (\ref{defgam}), we find for 
the rotational deformation 
\begin{equation}\label{rotdef}
\kappa_\alpha{}^\beta = {\stackrel \circ D}\omega^\beta{}_\alpha. 
\end{equation}

We thus recovered the deformation measures (\ref{Coss1}),(\ref{Coss2})
of the linear Cosserat continuum.  Using local coordinates, we
expand ${\stackrel \circ \vartheta}{}^\alpha = {\stackrel \circ
  e}{}_i{}^\alpha dx^i$, and then (\ref{trdef}) and (\ref{rotdef}) reduce
in tensor components to
\begin{eqnarray}
\beta_i{}^j &=& {\stackrel \circ \nabla}_i u^j - \omega^j{}_i,\\
\kappa_{ij}{}^k &=& {\stackrel \circ \nabla}_i\omega^k{}_j.
\end{eqnarray}
Thus, the deformation measures of the Cosserat continuum are literally
given by the deformations of coframe and connection (\ref{defvta})-(\ref{defgam}).

The compatibility conditions (\ref{compat}) can be derived {}from
(\ref{trdef}) and (\ref{rotdef}) by applying the covariant derivative.
The result reads
\begin{equation}
{\stackrel \circ D}\beta^\alpha + \kappa_\beta{}^\alpha\wedge 
{\stackrel \circ \vartheta}{}^\beta = 0,\qquad {\stackrel \circ D}
\kappa_\beta{}^\alpha = 0.\label{compati}
\end{equation}
The crucial point is that the geometry of the space is Euclidean and flat. 

When, however, the space has a nontrivial Riemann-Cartan geometry with the 
coframe $\vartheta^\alpha$ and connection $\Gamma_\alpha{}^\beta$ satisfy
Cartan's structure equations with the nontrivial torsion $T^\alpha$ and
curvature $R_\alpha{}^\beta$, the deformation measures are given by 
\begin{eqnarray}\label{meas1}
\beta^\alpha &=& Du^\alpha - \omega^\alpha{}_\beta\,\vartheta^\beta + 
u\rfloor T^\alpha,\\\label{meas2}
\kappa_\alpha{}^\beta &=& D\omega^\beta{}_\alpha + u\rfloor R_\alpha{}^\beta,
\end{eqnarray}
and they no longer satisfy the compatibility conditions
(\ref{compati}). In 4D, after suitably adjusting the signs,
Eqs.(\ref{meas1}) and (\ref{meas2}) coincide with the Poincar\'e gauge
transformations (\ref{PG1}), (\ref{PG2}).


\vskip 30pt
\begin{eref}

\bibitem{AldrovandiPereira} R.~Aldrovandi and J.G.~Pereira, {\it An
    introduction to teleparallel gravity}, (2001-2007) Lecture notes,
  unpublished (112 pages), available {}from
  http://www. ift.unesp.br/users/jpereira/classnotes.html

\bibitem{AndradePereira} V.C.~de Andrade and J.G.~Pereira, {\it
    Gravitational Lorentz force and the description of the
    gravitational interaction}, {Phys. Rev.} {\bf D56} (1997)
  4689--4695. 

\bibitem{Badur} J.~Badur and H.~Stumpf, {\it On the influence of E.\
    and F.~Cosserat on modern continuum mechanics and field theory,}
  University of Bochum, Institute for Mechanics, Communication number
  {\bf 72} (December 1989) 39 pages.

\bibitem{Becker} R.E.~Becker, {\it Letter,} Physics Today {\bf 59} (April
  2006) 14--15.

\bibitem{Bilby} B.A.~Bilby, R.~Bullough and E.~Smith, {\it Continuous
    distributions of dislocations: a new application of the methods of
    Non-Riemannian geometry,} Proc.\ Roy.\ Soc. (London) {\bf A 231}
  (1955) 263--273.

\bibitem{Milutin} M.~Blagojevi\'c, {\it Gravitation and Gauge
    Symmetries,} IoP, Bristol, UK (2002).

\bibitem{MilutinTorsion} M.~Blagojevi\'c and B.~Cvetkovi\'c, {\it
    Black hole entropy from the boundary conformal structure in 3D
    gravity with torsion,} JHEP {\bf 0610} (2006) 005 (12 pages)
  [arXiv:gr-qc/0606086].

\bibitem{Braaten} E.~Braaten, T.L.~Curtright, and C.K.~Zachos, {\it
    Torsion and geometrostasis in nonlinear sigma models}, {Nucl.\
    Phys.} {\bf B260} (1985) 630--688.

\bibitem{Lasenby} S.D.~Brechet, M.P.~Hobson, A.N.~Lasenby, {\it
    Weyssenhoff fluid dynamics in general relativity using a 1+3
    covariant approach}, Class.\ Quant.\ Gravity, to be published
  (Dec.\ 2007).

\bibitem{Capriz} G.~Capriz, {\it Continua with Microstructure},
Springer Tracts Nat. Phil. Vol. {\bf35} (1989)

\bibitem{Carroll} S.M.~Carroll, {\it Spacetime and Geometry,} An
  Introduction to General Relativity, Addison Wesley, San Francisco
  (2004).

\bibitem{Cartan1922} \'E.~Cartan, {\it Sur une g\'en\'eralisation de
    la notion de courbure de Riemann et les espaces \`a torsion},
  {C.R.\ Acad.\ Sci. (Paris)} {\bf 174} (1922) 593--595; English
  translation by G.D.~Kerlick: {\it On a generalization of the notion
    of Riemann curvature and spaces with torsion}, in: {\sl ``Proc. of
    the 6th Course of Internat. School on Cosmology and Gravitation:
    Spin, Torsion, Rotation, and Supergravity" (Erice, 1979) Eds.
    P.G.Bergmann and V.De Sabbata} (Plenum: New York, 1980) 489--491;
  with subsequent comments of A.~Trautman, {\it Comments on the paper
    by \'Elie Cartan: Sur une g\'en\'eralisation de la notion de
    courbure de Riemann et les espaces \`a torsion}, pp.\ 493--496.

\bibitem{Cartan1986} \'E. Cartan, {\it On Manifolds with an Affine
    Connection and the Theory of General Relativity}, English transl.\
  of the French original by A.~Magnon and A.~Ashtekar, Bibliopolis,
  Napoli (1986).

\bibitem{Cartan2001}\'E.~Cartan, {\it Riemannian Geometry in an
    Orthogonal Frame}, {}from lectures delivered in 1926/7, transl. {}from
  the Russian by V.V. Goldberg, World Scientific, New Jersey (2001).

\bibitem{Cho1976} Y.M.~Cho, {\it Einstein Lagrangian as the
    translational Yang-Mills Lagrangian,} Phys.\ Rev. {\bf D14}
  (1976) 2521--2525.

\bibitem{Cosserat} E. et F.~Cosserat, {\it Th\'eorie des corps
    d\'eformables}, Hermann, Paris (1909), translated into English by
  D.~Delphenich (2007).

\bibitem{Costa} O.~Costa de Beauregard, {\it Translational inertial
    spin effect}, {Phys. Rev.} {\bf 129} (1963) 466--471.
  
\bibitem{Dereli} T.~Dereli and R.W.~Tucker, {\it An Einstein-Hilbert
    action for axi-dilaton gravity in four dimensions}, {Class.
    Quantum Grav.} {\bf 12} (1995) L31--L36.

\bibitem{Evans2003} M.W.~Evans, {\it A generally covariant field
    equation for gravitation and electromagnetism}, Foundations of
  Physics Letters {\bf 16} (2003) 369--377.

\bibitem{Evans2005a} M.W.~Evans, {\it The spinning and curving of
spacetime: The electromagnetic and gravitational fields in the Evans
field theory,} Foundations of Physics Letters {\bf 18} (2005)
431--454.

\bibitem{Frankel} T.~Frankel, {\it The Geometry of Physics}, 2nd ed.,
  Cambridge, UK (2004).

\bibitem{Garcia} A.A.~Garc{\'\i}a, F.W.~Hehl, C.~Heinicke and
  A.~Mac{\'\i}as, {\it Exact vacuum solution of a (1+2)-dimensional
    Poincar\'e gauge theory: BTZ solution with torsion,} Phys.\ Rev.\
  D {\bf 67} (2003) 124016 (7 Pages) [arXiv:gr-qc/0302097].

\bibitem{GronwaldIJMPD} F.~Gronwald, {\it Metric-affine gauge theory
    of gravity: I. fundamental structure and field equations}, {Int.\
    J.\ Mod.\ Phys.} {\bf D6} (1997) 263--303.

\bibitem{GronHehl} F. Gronwald and F.W. Hehl: {\it Stress and
    hyperstress as fundamental concepts in continuum mechanics and in
    relativistic field theory.} In: `Advances in Modern Continuum
  Dynamics', International Conference in Memory of Antonio Signorini,
  Isola d'Elba, June 1991.  G. Ferrarese, ed. (Pitagora Editrice,
  Bologna, 1993) pp.\ 1--32; arXiv.org/abs/gr-qc/9701054.

\bibitem{Erice} F.~Gronwald and F.W.~Hehl, {\it On the gauge aspects
    of gravity,} in {\sl Proc.\ Int.\ School of Cosm.\ \& Gravit.}
  14\({}^{{\rm th}}\) Course: Quantum Gravity. Held in Erice, Italy.
  Proceedings, P.G.\ Bergmann et al.\ (eds.). World Scientific,
  Singapore (1996) pp.\ 148--198; http://arxiv.org/abs/gr-qc/9602013.

\bibitem{Guenther} W.~G\"unther, {\it Zur Statik und Kinematik des
    Cosseratschen Kontinuums,} Abh.\ Braunschw.\ Wiss.\ Ges. {\bf 10}
    (1958) 195--213.

\bibitem{Hammond} R.T.~Hammond: {\it New fields in general relativity.}
  Contemporary Physics {\bf 36} (1995) 103--114.

\bibitem{HansonRegge} A.J.~Hanson and T.~Regge, {\it Torsion and
    quantum gravity}, in Lecture Notes in Physics (Springer) {\bf 94}
  (1979) 354--360.

\bibitem{Hartley} D.~Hartley, {\it Normal frames for non-Riemannian
    connections}, Class.\ Quantum Grav. {\bf 12} (1995) L103--L105.

\bibitem{HSh79} K. Hayashi and T. Shirafuji, {\it New general
    relativity}, {Phys. Rev.} {\bf D19} (1979) 3524--3553.

\bibitem{College} F.W.~Hehl, {\it Fermions and Gravity.} Colloque du
  Centenaire de la Naissance d'Albert Einstein au Coll\`ege de France,
  June 1979, pp.119--148. Edition du Centre National Recherche
  Scientifique, Paris (1980).

\bibitem{Hehl1997} F.W.~Hehl, {\it Alternative gravitational theories
    in four dimensions,} in: Proc.\ 8th M. Grossmann Meeting,
  T.~Piran, ed., World Scientific, Singapore (1998);
  arXiv:gr-qc/9712096 (10 pages).

\bibitem{AssessI} F.W.~Hehl, {\it An assessment of Evans' unified
    field theory I,} Foundations of Physics, to be published (2007/8);
  arXiv:physics/0703116 (36 pages).

\bibitem{HvdH} F.W.~Hehl and P.~von~der~Heyde, {\it Spin and the
    structure of spacetime,} Ann.\ Inst.\ H.\ Poincar\'e {\bf A19}
  (1973) 179--196.

\bibitem{RMP} F.W. Hehl, P. von der Heyde, G.D. Kerlick, and J.M.
  Nester, {\it General relativity with spin and torsion: Foundations
    and prospects}, Rev.\ Mod.\ Phys.  {\bf 48} (1976) 393--416.

\bibitem{HK}F.W.~Hehl and E.~Kr\"oner, {\it On the constitutive law of
    an elastic medium with moment stresses (in German),} Z.\ f.\
  Naturf. {\bf 20a} (1965) 336--350.

\bibitem{HMcCrea} F.W. Hehl and J.D. McCrea, {\it Bianchi identities
    and the automatic conservation of energy--momentum and angular
    momentum in general-relativistic field theories,} Foundations of
  Physics {\bf 16} (1986) 267--293.

\bibitem{PRs} F.W. Hehl, J.D. McCrea, E.W.  Mielke, and Y. Ne'eman:
  {\it Metric-affine gauge theory of gravity: Field equations, Noether
    identities, world spinors, and breaking of dilation invariance}.
  Phys. Rep. {\bf 258} (1995) 1--171.

\bibitem{birkbook} F.W.~Hehl and Yu.N.~Obukhov, {\it Foundations of
    Classical Electrodynamics: Charge, Flux, and Metric}
  (Birkh\"auser: Boston, MA, 2003).

\bibitem{AssessII} F.W.~Hehl and Yu.N.~Obukhov, {\it An assessment of
  Evans' unified field theory II,} Foundations of Physics, in press
  (2007/8); arXiv:physics/0703117 (11 pages).

\bibitem{Itin0} Y.~Itin, {\it Energy-momentum current for coframe
    gravity,} Class.\ Quant.\ Grav. {\bf 19} (2002) 173--190
  [arXiv:gr-qc/0111036].

\bibitem{Itin} Y.~Itin, {\it Noether currents and charges for
    Maxwell-like Lagrangians}, {J.\ Phys.\ A: Math. Gen.} {\bf 36}
  (2003) 8867--8883.

\bibitem{Itin2} Y.~Itin, {\it Weak field reduction in teleparallel
    coframe gravity: Vacuum case,} J.\ Math.\ Phys. {\bf 46} (2005)
    012501 (14 pages) [arXiv:gr-qc/0409021].

  \bibitem{Kaempfer} F.A. Kaempfer, {\it Vierbein field theory of
      gravity}, {Phys.\ Rev.} {\bf 165} (1968) 1420-1423.

\bibitem{Katanaev} M.~O.~Katanaev, {\it Geometric Theory of Defects,}
    Phys.\ Usp. {\bf 48} (2005) 675--701 [Usp.\ Fiz.\ Nauk {\bf 175}
    (2005) 705--733] [arXiv:cond-mat/0407469].

  \bibitem{Kibble} T.W.B. Kibble, {\it Lorentz invariance and the
      gravitational field}, J.\ Math.\ Phys. {\bf 2} (1961) 212--221.

\bibitem{Kiehn} R.M.~Kiehn, {\it The many faces of torsion,} Preprint
  (22 pages) September 2007.

\bibitem{Kleinert} H.~Kleinert and S.V.~Shabanov, {\it Spaces with
    torsion {}from embedding, and the special role of autoparallel
    trajectories}, {Phys.\ Lett.} {\bf B428} (1998) 315--321.

\bibitem{Kondo1952} K.~Kondo, {\it On the geometrical and physical
    foundations of the theory of yielding,} in: {\sl Proc. 2nd Japan
    Nat.\ Congr.\ Applied Mechanics, Tokyo}, pp.41--47 (1952).

\bibitem{Kroener1958} E.~Kr\"oner, {\it Kontinuumstheorie der
Versetzungen und Eigenspannungen,}
Erg.\ Ang.\ Math.\ (Springer) Vol. {\bf 5} (1958).

\bibitem{Kroener1980} E.~Kr\"oner, {\em Continuum theory of defects},
  in: {\sl Physics of Defects}, Les Houches, Session XXXV, 1980,
  R.Balian et al., eds., North-Holland, Amsterdam (1981) p.\ 215.

\bibitem{Kroener1986} E.~Kr\"oner, {\it The continuized crystal --- a
  bridge between micro-- and macromechanics}, Z.\ angew.\ Math.\ Mech.
  (ZAMM) {\bf 66} (1986) T284.

\bibitem{Kroener1993} E.~Kr\"oner, {\it A variational principle in
    nonlinear dislocation theory,} in: Proc.\ 2nd Int.\ Conf.\
  Nonlin.\ Mechanics, Chien Wei-zang, ed., Peking University Press,
  Beijing (1993) pp.59--64.

\bibitem{Lazar} M.~Lazar, {\it Dislocation theory as a 3-dimensional
    translation gauge theory}, Ann.\ Phys. (Leipzig) {\bf 9} (2000)
  461--473; arxiv.org/abs/cond-mat/0006280.

\bibitem{LazarA} M.~Lazar and C.~Anastassiadis, {\em The gauge theory
    of dislocations: static solutions of screw and edge dislocations,}
  preprint, TU Darmstadt, October 2007.  (38 pages).

\bibitem{Maluf} J.W.~Maluf and J.F.~da Rocha-Neto, {\it Hamiltonian
    formulation of general relativity in the teleparallel geometry},
  {Phys.\ Rev.} {\bf D64} (2001) 084014 (8 pages).

\bibitem{Malyshev} C.~Malyshev, {\it The Einsteinian T(3)-gauge
    approach and the stress tensor of the screw dislocation in the
    second order: avoiding the cut-off at the core}, J.\ Phys. {\bf A
    40} (2007) 10657--10684.

\bibitem{Sardan} L.~Mangiarotti and G.~Sardanashvily, {\it Connections
    in Classical and Quantum Field Theory}, World Scientific,
  Singapore (2000).

\bibitem{Mao} Y.~Mao, M.~Tegmark, A.~Guth and S.~Cabi, {\it
    Constraining torsion with Gravity Probe B,} Phys.\ Rev. {\bf D}
  (2007/8) to be published, arXiv:gr-qc/0608121.

\bibitem{McMap} J.D.~McCrea, F.W.~Hehl, and E.W.~Mielke, {\it Mapping
    Noether identities into Bianchi identities in general relativistic
    field theories of gravity and in the field theory of static
    lattice defects,} Int.\ J.\ Theor.\ Phys. {\bf 29} (1990)
  1185--1206.

\bibitem{Meyer} H.~Meyer, {\it Moller's tetrad theory of gravitation
    as a special case of Poincar\'e theory -- a coincidence?}, {Gen.\
    Relat.\ Grav.} {\bf 14} (1982) 531--548.

\bibitem{Mindlin} R.D.~Mindlin, {\it Micro-structure in linear
    elasticity,} Arch.\ Rat.\ Mech.\ Anal. {\bf 16} (1964) 51--78.

\bibitem{Minkevich} A.V.~Minkevich, {\it Gravitation, cosmology and
    space-time torsion}, arxiv.org/abs/ 0709.4337 (10 pages).
 
%
\bibitem{Moller} C.~M\o{}ller, {\it Conservation laws and absolute
    parallelism in general relativity}, {Mat.\ Fys.\ Skr.\ Dan.\ Vid.\
    Selsk.} {\bf 1}, no.\ 10 (1961) 1--50.

\bibitem{Muench97} U. Muench, {\it \"Uber teleparallele
    Gravitationstheorien}, Diploma Thesis, University of Cologne
  (1997).

\bibitem{Nabarro} F.R.N.~Nabarro, {\it Theory on Crystal
    Dislocations}, Dover, N.Y. (1987).

 \bibitem{NH} Y.~Ne'eman and F.W.~Hehl, {\it Test matter in a
    spacetime with nonmetricity,} Class.\ Quant.\ Grav. {\bf 14}
  (1997) A251--A260 [arXiv:gr-qc/9604047].

\bibitem{Jim} J.~M.~Nester, {\it Gravity, torsion and gauge theory,}
  in: {\it Introduction to Kaluza--Klein theories}, H.C. Lee, ed.
  (World Scientific, Singapore (1984) pp.83--115.

\bibitem{Nitsch1979} J.~Nitsch and F.~W.~Hehl, {\it Translational
      gauge theory of gravity: post-Newtonian approximation and spin
      precession,} Phys.\ Lett. {\bf B90} (1980) 98--102.

\bibitem{Nye1953} J.F.~Nye, {\it Some geometrical relations in
    dislocated crystals,} Acta metallurgica {\bf 1} (1953) 153--162.

\bibitem{Yuridiffgeo} Yu.N.~Obukhov, {\it Poincar\'e gauge gravity:
    Selected topics,} Int.\ J.\ Geom.\ Meth.\ Mod.\ Phys. {\bf 3}
  (2006) 95--138; [arXiv:gr-qc/0601090].
 
\bibitem{OP} Yu.N.~Obukhov and J.G.~Pereira, {\it Metric-affine
    approach to teleparallel gravity}, {Phys.\ Rev.} {\bf D67}
  (2003) 044016 (17 pages).

\bibitem{Ohanian} H.C.~Ohanian and R.~Ruffini, {\it Gravitation and
    Spacetime,} 2nd ed., Norton, New York (1994).

\bibitem{Pele} C. Pellegrini and J. Plebanski, {\it Tetrad fields and
    gravitational fields}, {Mat.\ Fys.\ Skr.\ Dan.\ Vid.\ Selsk.} {\bf
    2}, no.\ 4 (1963) 1--39.

\bibitem{Pol2} J.~Polchinski, {\it String Theory Volume II,
    Superstring Theory and Beyond,} Cambridge University Press,
  Cambridge, UK (1998).

\bibitem{Pono} V.N.\ Ponomariov and Yu.\ Obukhov: {\it The generalized
    Einstein-Maxwell theory of gravitation.} Gen. Relat.  Grav. {\bf
    14} (1982) 309--330.

\bibitem{Sazdovic} D.S.~Popovi\'c and B.~Sazdovi\'c, {\it The
    geometrical form for the string space-time action}, {Eur.\ Phys.\
    J.} {\bf C50} (2007) 683--689.

\bibitem{Puetzfeld} D.~Puetzfeld, {\it Status of non-Riemannian
    cosmology}, {\sl New Astronomy Reviews} {\bf 49} (2005) 59-64.

\bibitem{POPRL} D.~Puetzfeld and Yu.N.~Obukhov, {\it Probing
    non-Riemannian spacetime geometry,} arXiv:0708.1926 [gr-qc] (4
    pages).
 
\bibitem{PO} D.~Puetzfeld and Yu.N.~Obukhov, {\it Propagation
    equations for deformable test bodies with microstructure in
    extended theories of gravity}, {Phys.\ Rev.} {\bf D76} (2007)
  084025 (20 pages).

\bibitem{Puntigam} R.A.~Puntigam and H.H.~Soleng, {\it Volterra
    distortions, spinning strings, and cosmic defects,} Class.\
  Quant.\ Grav. {\bf 14} (1997) 1129--1149 [arXiv:gr-qc/9604057].
 
\bibitem{RT} M.L.~Ruggiero and A.~Tartaglia, {\it Einstein-Cartan
    theory as a theory of defects in space-time,} Am.\ J.\ Phys. {\bf
    71} (2003) 1303.

\bibitem{Lewis} L.H.~Ryder and I.L.~Shapiro, {\it On the interaction
    of massive spinor particles with external electromagnetic and
    torsion fields,} Phys.\ Lett. {\bf A247} (1998) 21--26.
  [arXiv:hep-th/9805138].

\bibitem{Saa} A.~Saa, {\it A geometrical action for dilaton gravity},
  {Class.\ Quantum Grav.} {\bf 12} (1995) L85--L88.

\bibitem{SchaeferZAMM} H.~Schaefer, {\it Das Cosserat-Kontinuum}, ZAMM
  {\bf 47} (1967) 485--498.

\bibitem{Schaefer} H.~Schaefer, {\it Die Motorfelder des
    dreidimensionalen Cosserat-Kontinuums im Kalk\"ul der
    Differentialformen,} Int.\ Centre for Mechanical Sciences (CISM),
  Udine, Italy, Courses and Lectures (L.~Sobrero, ed.) No.\ {\bf 19}
  (60 pages) (1970).

\bibitem{ScherkSchwarz} J.~Scherk and J.H.~Schwarz, {\it Dual models
    and the geometry of space-time,} Phys.\ Lett. {\bf B52} (1974)
  347--350.

\bibitem{Schouten1954} J.A.~Schouten, {\it Ricci-Calculus}, 2nd ed.,
  Springer, Berlin (1954).

\bibitem{Schroedinger} E. Schr\"odinger: {\it Space--Time Structure},
  reprinted with corrections, Cambridge University Press, Cambridge
  (1960).

\bibitem{Schuecking} E.L.~Schucking, {\it The homogeneous
    gravitational field,} Foundations of Physics {\bf 15} (1985)
  571--577.

\bibitem{SchueckingSurowitz} E.L.~Schucking and E.J.~Surowitz, {\it
    Einstein's apple: his first principle of equivalence},
  arXiv:gr-qc/0703149 (30 pages).

\bibitem{Schweizer} M.~Schweizer, N.~Straumann and A.~Wipf, {\it
    Postnewtonian generation of gravitational waves in a theory of
    gravity with torsion,} Gen.\ Rel.\ Grav. {\bf 12} (1980) 951--961.

\bibitem{Sciama} D.W.~Sciama, {\it On the analogy between charge and spin
  in general relativity}, in: {\it Recent Developments of General
    Relativity}, Pergamon, London (1962) p.415.

\bibitem{Shapiro} I.~L.~Shapiro, {\it Physical aspects of the
    space-time torsion,} Phys.\ Rept.\ {\bf 357} (2002) 113--213
  [arXiv:hep-th/0103093].
  
\bibitem{Sharpe} R.W. Sharpe, {\it Differential Geometry: Cartan's
    Generalization of Klein's Erlangen Program,} Springer, New York
  (1997).

\bibitem{Simpson} J.H.~Simpson, {\it Nuclear gyroscopes,} Astronautics
  \& Aeronautics, October 1964, pp.42--48.

\bibitem{TrautmanHeld} A.~Trautman, {\it Fiber bundles, gauge fields
    and gravitation}, in: {\sl ``General Relativity and Gravitation:
    One Hundred Years after the Birth of Albert Einstein" A.Held, ed.}
  (Plenum: New York, 1980) vol. {\bf 1}, 287--308.

\bibitem{TrautmanSUNY} A.~Trautman, {\it Differential Geometry for
    Physicists, Stony Brook Lectures} Bibliopolis, Napoli (1984).

\bibitem{Trautman} A.~Trautman, {\it Einstein-Cartan theory,} in {\sl
    Encyclopedia of Math.\ Physics,} J.-P.\ Francoise et al., eds.,
  Elsevier, Oxford (2006) pp.\ 189--195; arXiv.org/gr-qc/0606062.

\bibitem{Tresguerres2007} R.~Tresguerres, {\it Translations and
    dynamics,} arXiv:0707.0296 [gr-qc] (21 pages).

\bibitem{TresguerresMielke2000} R.~Tresguerres and E.W.~Mielke, {\it
    Gravitational Goldstone fields {}from affine gauge theory,} Phys.\
  Rev.\ D {\bf 62} (2000) 044004 (7 pages); [arXiv:gr-qc/0007072].

\bibitem{Weertman} J. and J.A. Weertman, {\it Elementary Dislocation
    Theory}, MacMillan, London (1969).

%
\bibitem{WeinbergEinstein} S.~Weinberg, {\it Einstein's mistakes,}
  {\bf 58} (November 2005) 31--35.

\bibitem{WeinbergPT} S.~Weinberg, {\it Weinberg replies,} Physics
  Today {\bf 59} (April 2006) 15--16 (last paragraph of the reply);
  F.W.~Hehl and S.~Weinberg, {\it Note on the torsion tensor,} Physics
  Today {\bf 60} (March 2007) 16.

\bibitem{Weitzenboeck} R.~Weitzenb\"ock, {\it Invarianten-Theorie},
  Noordhoff, Groningen (1923) 416 pp.

\bibitem{Wise} D.K.~Wise, {\it MacDowell-Mansouri gravity and Cartan
    geometry,} arXiv.org/gr-qc/ 0611154.

\bibitem{YS} P.B.~Yasskin and W.R.~Stoeger, {\it Propagation equations
    for test bodies with spin and rotation in theories of gravity with
    torsion}, {Phys.\ Rev.} {\bf D21} (1980) 2081--2094.

%
\bibitem{grains} A.~Zeghadi, S.~Forest, A.-F.~Gourgues, O.~Bouaziz,
  {\it Cosserat continuum modelling of grain size effects in metal
    polycrystals,} PAMM -- Proc.\ Appl.\ Math.\ Mech. {\bf 5} (2005)
  79--82.

\end{eref}
\man{10 novembre 2007}

\end{document}